\documentclass[aip,jcp,amsmath,amssymb,reprint,longbibliography,groupedaddress]{revtex4-1}

\usepackage[utf8]{inputenc}
\usepackage{braket}
\usepackage{amsmath}
\usepackage{bbm}
\usepackage{appendix}
\usepackage{amsfonts}
\usepackage{comment}
\usepackage{cancel}
\usepackage{gensymb}
\usepackage{bbold}
\usepackage{graphicx}
\usepackage{MnSymbol}
\usepackage{mathtools}
\usepackage{tabularx}
\usepackage{array}
\usepackage{accents}
\usepackage{bm}
\usepackage{xcolor}
\usepackage{placeins}
\usepackage{enumitem}   
\usepackage{amsthm}
\usepackage{soul}
\usepackage[justification=Justified,singlelinecheck=off]{caption}

\newcommand{\bigbox}{\mathop{\raisebox{-0.5ex}{\scalebox{1.5}{$\square$}}}\displaylimits}

\newcolumntype{Y}{>{\centering\arraybackslash}X}

\usepackage[colorlinks, linkcolor=blue]{hyperref}
\usepackage[nameinlink, capitalise]{cleveref}

\newcommand{\nn}{\nonumber} 
\newcommand\numeq[1]%
{\stackrel{\scriptscriptstyle(\mkern-1.5mu#1\mkern-1.5mu)}{=}}

\newcommand{\tcb}{\textcolor{blue}}
\newcommand{\tcr}{\textcolor{red}}
\newcommand{\tcg}{\textcolor{green}}

\newcommand{\bs}{\boldsymbol}

\newcommand{\diag}{{\mathrm{diag}}}

\newcommand\commin[1]{\iffalse #1 \fi}

\newcommand{\mr}{\mathrm}

\newcommand{\mc}{\mathcal}

\newcommand{\QL}{\ensuremath{\mathrm{QL}}}
\newcommand{\NQL}{\ensuremath{{N_\mathrm{QL}}}}
\newcommand{\Ng}{\ensuremath{{N_G}}}

\newcommand{\e}{\ensuremath{\,\mathrm{e}}}

\renewcommand{\Re}{\ensuremath{\mathrm{Re}}}
\renewcommand{\Im}{\ensuremath{\mathrm{Im}}}
\newcommand{\da}{\ensuremath{\downarrow}}
\newcommand{\ua}{\ensuremath{\uparrow}}

\usepackage{chngcntr}

\newcommand\norm[1]{\left\lVert#1\right\rVert}

\numberwithin{equation}{section}

\newtheorem{axiom}{Axiom}

\usepackage{etoolbox}
\makeatletter
\renewrobustcmd{\cref}{\@osmcref{cref}}
\renewrobustcmd{\Cref}{\@osmcref{Cref}}
\def\@osmcref#1#2{%
	\begingroup
	\ifcsundef{r@#2}
	{}
	{\expandafter\expandafter\expandafter\expandafter\expandafter
		\expandafter\expandafter\def
		\expandafter\expandafter\expandafter\expandafter\expandafter
		\expandafter\expandafter\@osmcref@name
		\expandafter\expandafter\expandafter\expandafter\expandafter
		\expandafter\expandafter{%
			\expandafter\expandafter\expandafter
			\@thirdoffive\csname r@#2\endcsname}}%
	\ifcsundef{r@#2@cref}
	{}
	{\cref@gettype{#2}{\@osmcref@type}}%
	\ifboolexpr{not test {\ifdefvoid{\@osmcref@name}}
		and (test {\ifdefstring{\@osmcref@type}{thm}}
		or test {\ifdefstring{\@osmcref@type}{lemma}})}
	{\nameref{#2} (\@cref{#1}{#2})}
	{\@cref{#1}{#2}}%
	\endgroup
}
\makeatother

\crefname{axioms}{axiom}{axioms}
\Crefname{axiom}{Axiom}{Axioms}

\begin{document}
\title{Quantum information with quantum-like bits}
\author{Graziano Amati, Gregory D. Scholes}
\affiliation{Department of Chemistry, Princeton University, Princeton, NJ 08544, USA.\looseness=-1}

\date{\today}

\begin{abstract}
In previous work we have proposed a construction of quantum-like bits that could endow a large synchronizing classical system, for example of oscillators, with quantum-like function that is not compromised by decoherence. In the present paper we investigate further this platform of quantum-like states. 
Firstly, we discuss a general protocol on how to construct classical synchronizing networks that allow for emergent states. 
We then study how gates can be implemented on those states.
This suggests the possibility of quantum-like information processing on a special class of many-body classical systems. 
Finally, we show that our approach allows for non-Kolmogorov interference, a feature that separates our model from a classical probabilistic system.	
This paper aims to explore the mathematical structure of quantum-like resources distilled from classical synchronizing systems, and shows how arbitrary gates can be implemented by manipulating many-body correlations. 

\end{abstract}

\maketitle

\section{Introduction}

The present noisy-intermediate scale quantum era of quantum devices is witnessing the successful implementation and test of diverse quantum algorithms on medium-size quantum hardware, from search on unsorted databases, \cite{zhang2021} to variational quantum algorithms, \cite{cerezo2021} to simulations of open quantum dynamics. \cite{sun2024}
However, hardware of at least hundreds of thousands of physical qubits, way beyond the current engineering capabilities, is needed to obtain practical quantum advantage.
This involves storing and processing the information volume required by the most challenging and relevant computational problems. \cite{beverland2022}
The critical issue limiting the scalability of quantum resources is the buildup of uncontrolled noise and dissipation, leading to loss of coherence and fidelity. \cite{preskill2018, lau2022,gill2022}

In parallel with quantum engineering advancements, ongoing research investigates whether classical computing resources can continue to play a key role in information processing.
For example, tensor network simulations have been shown to outperform the most advanced quantum processors in selected algorithms. \cite{tindall2024} 
Efficient quantum-inspired protocols have been successfully developed, leveraging advancements in information processing derived from quantum logic.   \cite{gharehchopogh2023}

\begin{widetext}
\begin{figure*}
\includegraphics[width=7in]{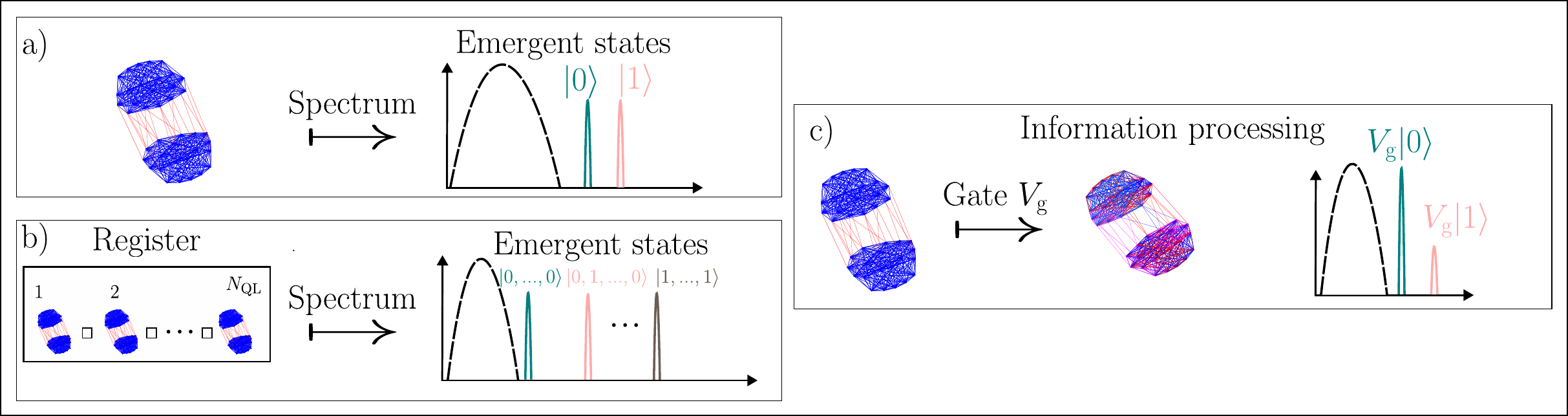}\caption{Illustration of the main concepts presented in this paper and in prior work by the authors. \cite{scholes2024,product} a) Many body systems consisting of networks of oscillators can easily exhibit classical synchronization phenomena. By structuring two-body correlations into two densely connected regions (blue) weakly interacting through sparse links (red), synchronization gives rise to pairs of emergent states. 
These states are well-distinguishable from a large band of random states at lower energies (dashed line in the sketch of the spectrum). In Ref.~\onlinecite{scholes2024} we proposed a map from emergent states to the computational basis of a qubit. b) In Ref.~\onlinecite{product}, we demonstrated that the emergent states of the Cartesian product of a set of synchronizing graphs scale with the same dimensionality of a qubit system. In this paper, we expand over these concepts by bridging classical synchronization dynamics and the QL formalism. c) Additionally, we elaborate on the concept of QL information processing, and how this connects to steady-state synchronization dynamics. We show that operations on graphs, implemented through suitable transformations of their edges, are equivalent to gate transformations $\hat V_{\mathrm g}$ applied directly to the emergent states. }\label{fig:illustration}
\end{figure*}
\end{widetext}

A large class of regular graphs encodes the correlations required to identify robust collective states.
This general feature can be exploited for a variety of applications. 
For example, we discussed in Ref.~\onlinecite{scholes2025excit} how this principle can 
support the engineering of molecular delocalization. 
The same reference, along with Refs.~\onlinecite{krebs2011,brouwer2011}, offers the interested reader background information on the mathematical formalism of regular and expander graphs, as well as their spectral properties.
In the spirit of quantum-inspired computing, we showed in recent work that network correlations can serve as resources for quantum information processing. \cite{scholes2020,scholes2023,scholes2024,product}
We focused, in particular, on many body systems whose correlations obey the graph structure shown in panel a) of \cref{fig:illustration}. Here, two densely connected regions (in blue) of a network are mutually interacting via sparse links (in red). A spectral analysis of this  class of graphs highlights the presence of two isolated emergent states.
These can be clearly distinguished from a broad band of states arising from random network configurations (dashed lines in the spectra of \cref{fig:illustration}). 
The robustness of emergent states against noise appears to be a direct consequence of synchronization. \cite{scholes2023,scholes2020}
This intriguing feature is related to the occurrence of classical synchronization phenomena. \cite{haken1975, fano1992,strogatz2012,artime2022}
We showed in Ref.~\onlinecite{scholes2024} that the emergent states of connected expander graphs are isomorphic to the Hilbert space of a qubit. 
This led us to introducing the notion of quantum-like (QL) bits within the context of classically-synchronizing networks. In Ref.~\onlinecite{product} we built on these results and proposed a general framework to construct resources for multiple QL bits. In particular, we showed that the Cartesian product ($\square$) of $\NQL$ bipartite graphs obeys the same dimensionality of a register of $\NQL$ qubits [see panel b) of \cref{fig:illustration}].

In the present paper, we build on the foundational work of Ref.~\onlinecite{scholes2024,product,scholes2023}, by
proposing a self-contained formalization of the concepts of QL states and gates and their application to classical many-body systems. 
As sketched in \cref{fig:illustration}, we present a framework to map the emergent states of a classical synchronizing network of oscillators onto a qubit Hilbert space. 
We highlight a connection between emergent states and steady-state synchronization dynamics, and we show that the latter is robust even in presence of stochastic disorder in the network.
This makes the emergent states viable candidates for QL information processing, which we implement by transforming the matrix of the two-body correlations between the oscillators.
Our analysis is general and applies to any classical system whose two-body correlations align with the network structures discussed later in this paper. 
However, to make the approach more concrete, we provide explicit examples in this manuscript by examining the synchronization dynamics of the Kuramoto model.
This is a non-symplectic classical system consisting of a network of nonlinear oscillators, whose collective properties have been extensively explored in the literature. \cite{kuramoto1975,kawamura2010,scholes2022,rodrigues2016}

Our approach leads to a natural correspondence between ``QL gates'' transformations and the steady states of the classical oscillatory dynamics.
Additionally, we propose a variety of experimental resources that can be utilized to test the formalism.
This supports the applicability of the approach for information protocols involving low-dimensional Hilbert spaces, where the exponentially large complexity of the classical phase space can be handled by available experimental resources.
Finally, we connect our perspective to the QL probabilistic framework developed by Khrennikov. \cite{khrennikov2007, khrennikov2019, khrennikov2010}
This theory formalizes the notion that genuine quantum systems are not required to process information in a QL way. 
Instead, to observe nonclassical interference it is sufficient to construct a contextual probability theory involving suitable notions of QL states, observables, and projective measurements.
This paper can be read alongside other recent developments by the authors within the same framework, as presented in the complementary work in Ref.~\onlinecite{encoding}. In that study, we show that arbitrary quantum states can be encoded into the synchronizing steady state of a nonlinear oscillator system. We also apply the approach proposed here to construct a QL analog of entanglement. Although the two works are closely related, each is self-contained and can be read independently.

The present paper is structured as follows.
In \cref{subsec:one} we summarize our previous work describing how to construct a single-QL-bit Hilbert space.
In \cref{subsec:many} we discuss its generalization to higher dimensions, and we show explicitly in \cref{subsubsec:bell} how to construct Bell states for two QL bits. 
In \cref{sec:sync_dyn} we show the fundamental role of emergent states in the synchronizing dynamics of the Kuramoto model as a paradigmatic classical system of synchronizing oscillators.
In \cref{sec:gates} we study the action of a complete set of QL gates, and we provide concrete examples of their action.
In \cref{sec:quantum_prob}, we present a set of axioms that enables us to define a QL probabilistic theory in our approach.
In particular, we highlight the presence of a nonclassical interference term in the relation between total and conditional probabilities, specific to non-Kolmogorovian contextual probabilistic models. \cite{khrennikov2010}
Finally, in \cref{sec:perspectives}, we propose a variety of experimental resources that could be utilized to experimentally implement and verify our formalism.

In conjunction with Refs.~\onlinecite{scholes2023, product, scholes2024}, this work provides the mathematical foundation for a rigorous approach to QL information processing from classical many-body synchronizing systems.

\section{Quantum-like bits}\label{sec:QLbits}

In this section we derive a mapping from the emergent states of classical networks to the Hilbert space of interacting QL bits. 
While the concept has been already elucidated in Ref.~\onlinecite{scholes2024}, here we lay out a more formal description. 
The present analysis provides a complete description of the axioms of states and operators in quantum mechanics, within the notation of the emergent states.
Later in the paper we show the direct connection between this abstract approach and the dynamics of classical nonlinear oscillators.

\subsection{Isolated system}\label{subsec:one}

Let us consider two classical many-body systems with pairwise interactions between the elementary constituents mapped onto two graphs, $G_1$ and $G_2$. 
Here, we assume here that both graphs involve the same number of vertices, $\Ng$, and we label by $A_1$ and $A_2$ the related adjacency matrices. 
Also, we consider at this stage undirected simple graphs, without loops (while in the paper we will discuss graph transformations that break these symmetries).
The adjacency matrix for this class of graphs is symmetric with entries either $0$ or $1$ and null diagonal.
Additionally, we assume an isotropic structure for all two-body interactions, such that the valence of the graphs is constant and equal to $k$. \cite{diestel2017}
In particular,
\begin{equation}\label{eq:ave_val}
	 \sum_{i=1}^\Ng [A_n]_{ij'} = \sum_{j=1}^\Ng [A_n]_{i'j} = k, \hspace{8mm} \forall i',j'=1,\ldots, \Ng, 
\end{equation}
and $n=1,2$. 
The direct sum of the two adjacency matrices is
\begin{equation}\label{eq:A12}
A_1 \oplus A_2 = \begin{bmatrix}
A_1 & \mathbb 0_\Ng \\
\mathbb 0_\Ng &  A_2
\end{bmatrix},
\end{equation}
where we labeled by $\mathbb 0_\Ng \in M_\Ng(\mathbb N)$ the null matrix on the off-diagonal blocks. 
As proposed in Ref.~\onlinecite{scholes2024}, we define a QL bit resource by switching on an interaction between the two subgraphs, and define the resource
\begin{equation}\label{eq:R}
\mc R = \begin{bmatrix}
A_{1} &  C \\
 C^\top &  A_{2}
\end{bmatrix},
\end{equation}
involving a coupling matrix $C$.
Note that $C$ does not represent a simple graph by itself, i.e. it is in general non-symmetric, and it can admit diagonal entries connecting corresponding vertices in $A_{1}$ and $A_{2}$. 
Conversely, \cref{eq:R}, being overall symmetric and loop-free, describes a simple undirected graph. 
The matrix $C$ is also chosen to be regular, with valency $l$.
\Cref{eq:R} represents the fundamental resource used to build an isolated QL bit. \cite{scholes2024} 

\begin{figure}
	\begin{minipage}{.45\linewidth}
		\centering
		\includegraphics[width=\linewidth]{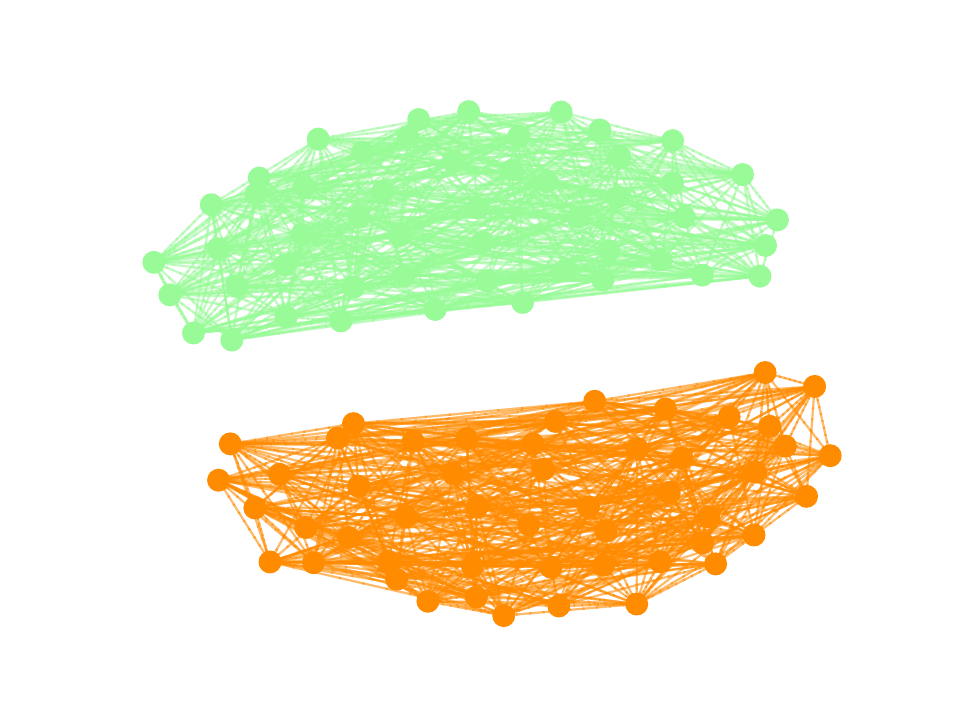}
		\caption*{\centering{$l=0$}}
		\label{fig:graph_l0}
	\end{minipage}
	\quad
\begin{minipage}{.45\linewidth}
	\centering
	\includegraphics[width=\linewidth]{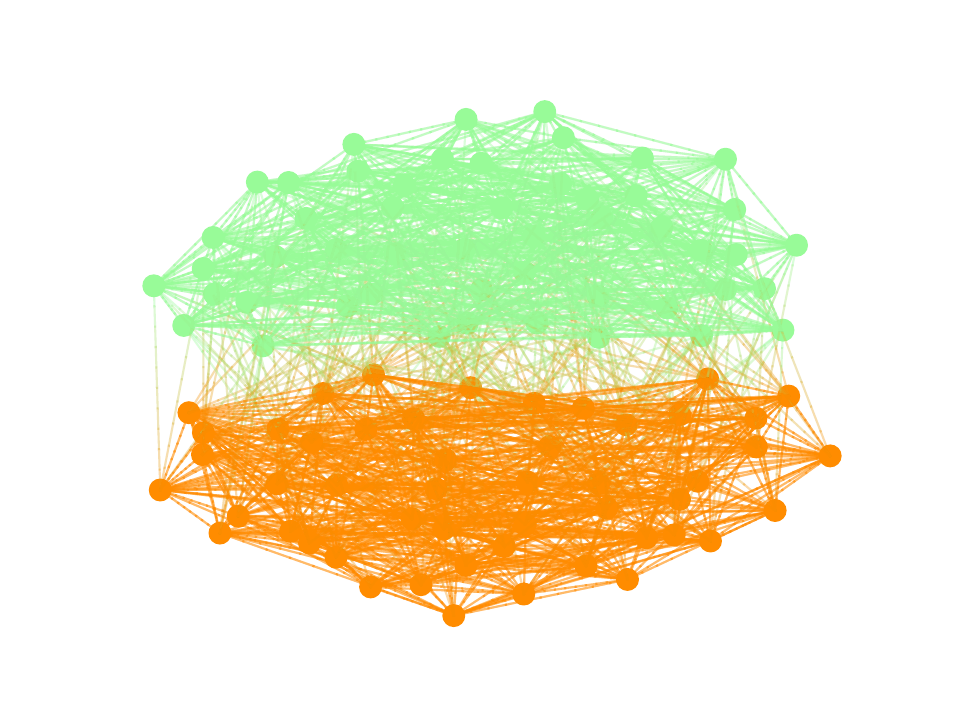}
	\caption*{\centering{$l=k/4$}}
\end{minipage}
	\begin{minipage}{.45\linewidth}
	\centering
	\includegraphics[width=\linewidth]{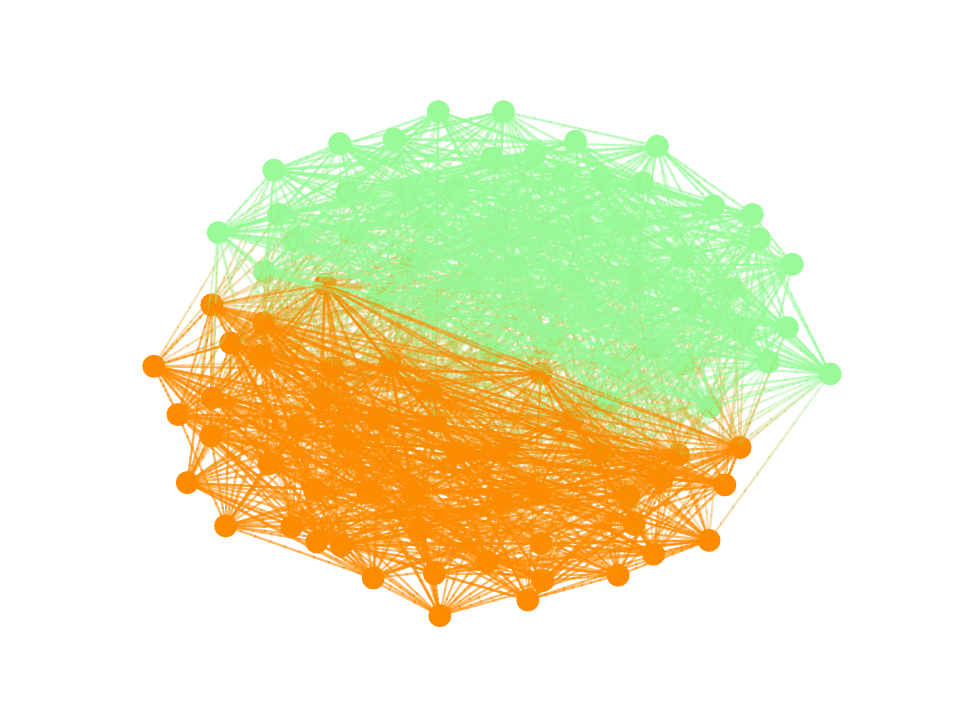}
	\caption*{\centering{$l=k/2$}}
\end{minipage}
	\quad
	\begin{minipage}{.45\linewidth}
	\centering
	\includegraphics[width=\linewidth]{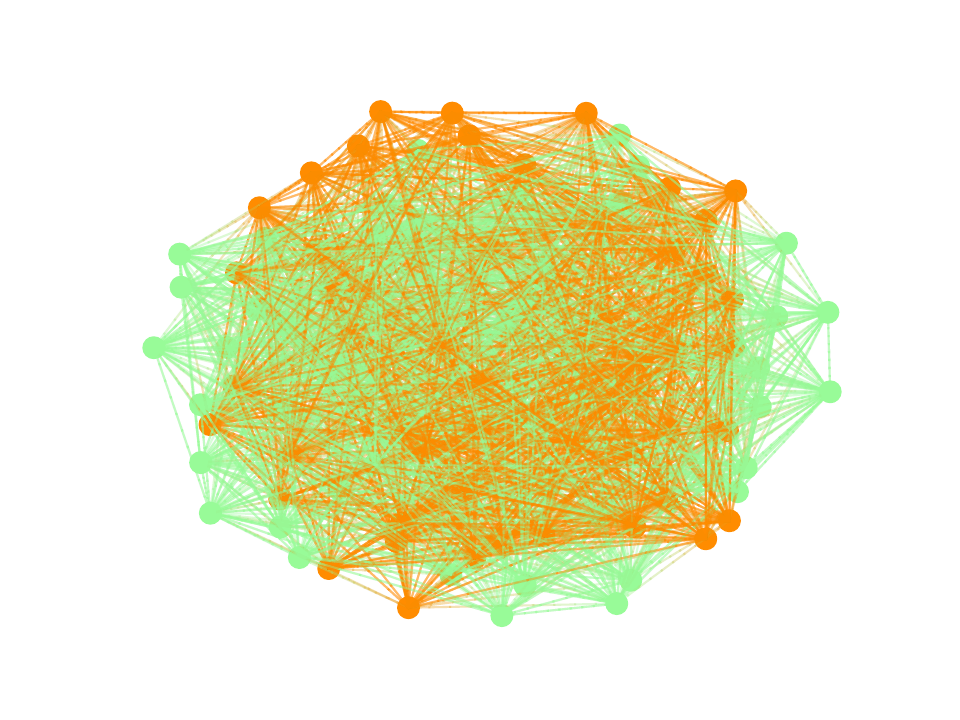}
	\caption*{\centering{$l=k$}}
\end{minipage}
	\caption{Graphs associated to the resource \cref{eq:R}, for increasing values of the valency $l$ of the coupling matrix $C$, as specified in the captions. $l$ is given here in units of the valency $k$ of the subgraphs $A_1$ (light green/light) and $A_2$ (dark orange/dark). The other parameters are fixed to $\Ng=50$ and $k=20$.}
	\label{fig:1QL}
\end{figure}

To better describe the structure of the graphs that are suitable resources for our QL platform, we show in \cref{fig:1QL} four instances of \cref{eq:R} for increasing values of $l$, as specified in the captions. 
The case $l=0$ corresponds to uncoupled regular graphs, as in \cref{eq:A12}. 
By increasing $l$, we strengthen the coupling between the two subgraphs.
For $l=k$, \cref{eq:R} turns into a regular matrix with total valency $2k$, with the same density of both intra- and cross-subgraphs edges.

As discussed in  Refs.~\onlinecite{scholes2023,scholes2024}, resources with structure as in \cref{eq:R} are particularly appealing for information processing.
In fact, these resources allow for emergent states associated to distinguishable eigenvalues, well isolated from a large band of random states.
The occurrence of emergent states is a common feature of  Erd\"os-Renyi random graphs and expander families, such as the k-regular graphs. \cite{scholes2020entropy, scholes2024,lubotzky2012}
By mapping the emergent states onto a Hilbert space, we obtain a platform for information processing that can be controlled by suitable network processing. 
We will elaborate on these important concepts further in the manuscript. 

\begin{figure}
\includegraphics[width=3.7in]{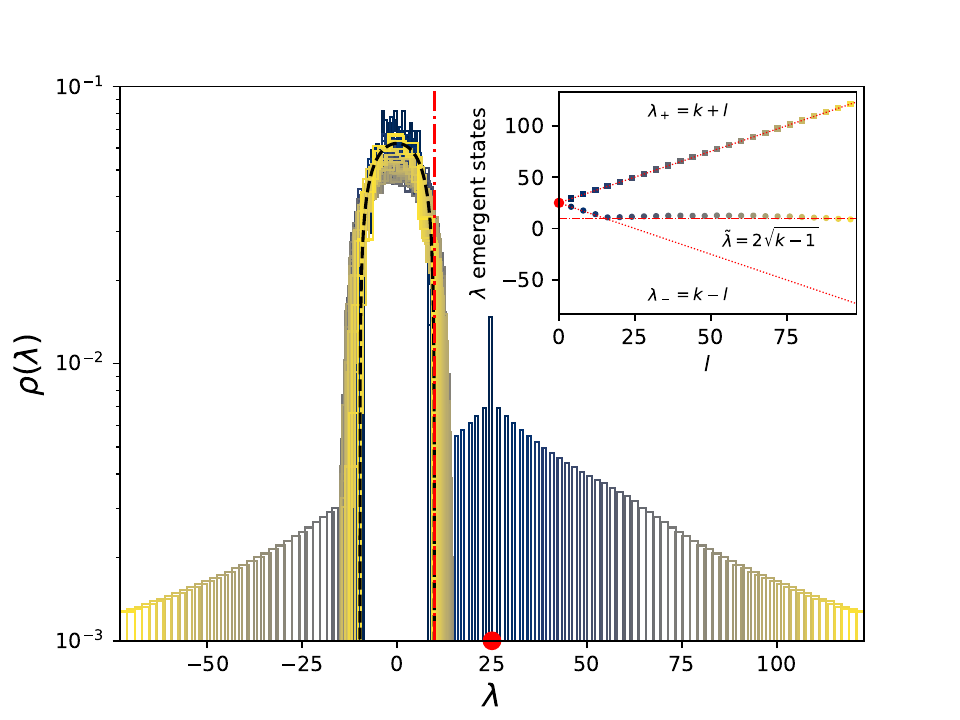}\caption{Main panel: distribution of the eigenvalues $\lambda$ of the QL resource \cref{eq:R}, for increasing values of the order $l$ of the coupling matrix $C$, from $l=0$ (blue/dark) to $l=\Ng$ (yellow/light). Inset: largest and second largest eigenvalues of the spectrum as a function of $l$, with the same color code as in the main panel. We fix here $\Ng=100$ and $k=25$. }\label{fig:histo_QL1}
\end{figure}

We study the spectral properties of the graph resources by plotting in \cref{fig:histo_QL1} the distribution $\rho(\lambda)$ of the eigenvalues of \cref{eq:R} for increasing values of the valency $l$ of the regular off-diagonal block $C$.
More specifically, we sample a set of graphs with the structure given in \cref{eq:R}. Given that each block of $\mc R$ defines a regular random graph, the only constraints given in the sampling are that each of the vertices in the $A_i$'s and $C$ is connected to $k$ and $l$ other vertices, respectively.
In \cref{eq:R} we show a family of distributions for a fixed value of $k=25$ and increasing vaues of $l$, from $l=0$ (blue/dark) to $l=\Ng$ (yellow/light). 
Here, the sizes of all blocks are fixed to $\Ng=100$.
The emergent states can be identified for each $l$ from the couples of isolated sharp peaks (with identical colors), displaced symmetrically around $\lambda=k=25$. 
More specifically, for $l=0$ we observe a single isolated line associated to the two-fold degenerate eigenvalue $\lambda=k=25$ (red markers). 
As the valency $l$ increases, this emergent eigenvalue split into two, and the peaks shift to
$\lambda_\pm = k\pm l$.
We track the ``ballistic'' evolutions of the eigenvectors as a function of the coupling valency $l$ in the inset of \cref{fig:histo_QL1}.
Here, the largest and second largest eigenvalues are shown as squared and dotted markers, respectively. 
While the value of the largest eigenvalue increases linearly for arbitrary $l$, we observe that the second largest eigenvalue plateaus at values close to $\tilde\lambda=2\sqrt{k-1}$.  
This barrier is defined by the radius of the Wigner semicircle distribution for a graph of order $k$, shown as a black dashed line in the figure. \cite{mckay1981,farkas2001}
This portion of the spectrum is populated by random low-frequency states, whose high density makes them hardly distinguishable. \cite{scholes2020,scholes2023,scholes2024} 
From the present analysis we infer that, by fixing $l = (k-\tilde\lambda)/2$, both the distance between the two emergent eigenvalues and between those and the dark states are maximized.
As elucidated in Ref.~\onlinecite{scholes2023,scholes2024}, this analysis  offers valuable insight for creating well-resolved and easily distinguishable emergent states for quantum information purposes.
\\

We discuss now how to construct the Hilbert space $\mc H^{(1)}$ of an isolated QL qubit from coupled regular graphs, with adjacency matrices with structure as in \cref{eq:R}.
In particular, we define the mapping
\begin{equation}\label{eq:psi_pm}
\psi_\da =  \tfrac 1 {\sqrt 2}\begin{pmatrix}
	a_1\\ 
	-a_2
\end{pmatrix}
\mapsto \ket {\da},  \hspace{6mm} 
\psi_\ua = \tfrac 1 {\sqrt 2}\begin{pmatrix}
	a_1\\ 
	a_2
\end{pmatrix}
\mapsto \ket {\ua}
\end{equation}
where $a_i$ is the normalized eigenvector associated to the largest eigenvalue of $A_i$.
As both the $A_i$'s and $C$ are exactly regular, then $a_i = (1,1\ldots, 1)^\top/\sqrt{2\Ng}$ are vectors with constant weights. 
Here and in the following we decided, however, to keep a more general notation, to account for possible structural disorder in the network that could break this symmetry.
The vectors $\psi_\sigma$, $\sigma \in \{\da,\ua\}\equiv \mc S$, are  eigenvectors of \cref{eq:R}  with eigenvalues $\{k- l,k+l\}$, respectively.
In earlier work, we used the QL bit construction to explicitly associate the subgraph basis with the emergent states: for example, the state $a_1$ of the graph with adjacency matrix $A_1$ represents $\ket\da$, and $a_2$ represents $\ket\ua$.\cite{scholes2024,product} \Cref{eq:psi_pm} presents a different, but equally valid, choice of basis.

The present approach establishes a direct connection between the computational basis and the emergent states of the composite graph described by \cref{eq:R}.
\\

In this section we summarized a list of prescriptions for defining a mapping from a bipartite graph to the Hilbert space of a single qubit, and related properties. \cite{product,scholes2024}
In the next \cref{subsec:many} we expand on an idea proposed in Ref.~\onlinecite{product}, to represent entanglement and superposition by coupling together multiple graph resources. 
Formal derivations on how to implement information processing and define a QL probabilistic model in this framework, follow in the later \cref{sec:gates,sec:quantum_prob}, respectively.

\subsection{Many-body system}\label{subsec:many}

\subsubsection{Formalism}\label{subsubsec:many}

Various representations of quantum states are possible with respect to classical network structures.
For a small number of QL bits the representation defined in Ref.~\onlinecite{scholes2024} is useful.
In that work the Bell states, involving two QL bits, are designed to give insight on proof-of-principle examples of QL entanglement.
In Ref.~\onlinecite{product} we propose a formal representation of states applicable to an arbitrary number of QL bits.
This is based on the Cartesian product of single-QL-bit graphs.
Consistently with that work, we define here as a simple generalization of \cref{eq:R} the resource for a system of $\NQL$ bits
\begin{equation}\label{eq:R_NQL}
\mc R = \sum_{q=1}^\NQL \left[\bigotimes_{p=1}^{q-1}\mathbb{1}_{2\Ng}\right]\otimes \mc R^{(q)}\otimes  \left[\bigotimes_{p'=q+1}^{\NQL}\mathbb{1}_{2\Ng}\right],
\end{equation}
where each $\mc R^{(q)}$ has the structure shown in \cref{eq:R}.
\Cref{eq:R_NQL} is the adjacency matrix of the Cartesian product graph \cite{product}
\begin{equation}\label{eq:G}
\mc G =  \bigbox_{q=1}^\NQL  \mc G^{(q)}.
\end{equation}	
Here, $\mc G^{(q)}$ is the single-QL-bit graph with adjacency matrix $\mc R^{(q)}$.
The tensors
\begin{align}\label{eq:psi_sigma} 
\psi_{\bs\sigma} = \bigotimes_{q=1}^{\NQL}  \psi_{\sigma^{(q)}}\mapsto \bigotimes_{q=1}^{\NQL}  \ket {\sigma^{(q)}}= \ket{\bm \sigma}
\end{align}
are eigenvectors of \cref{eq:R_NQL}.
Different basis elements are labeled using the multi-index $\bm \sigma = \left\{ \sigma^{(1)},\ldots, \sigma^{(N_\QL)} \right\}\in \bs{\mc S}$, where $\bs {\mc S} = \mc S^\NQL $.
As discussed after \cref{eq:psi_pm}, each $\psi_{\sigma^{(q)}}$, with $\sigma^{(q)}\in\mc S$, is associated to one of the eigenvalues $k\pm l$ of $\mc R^{(q)}$. 
Let us label this eigenvalue $\lambda_{\sigma^{(q)}}$.
The action of $\mc R$ on $\psi_{\bs\sigma}$ is then given by %
\begin{align}\label{eq:R_psi_sigma}
\mc R\psi_{\bs \sigma} &= \sum_{q=1}^\NQL \left[\bigotimes_{p=1}^{q-1}\psi_{\sigma^{(p)}}\right] \otimes \left[\mc R^{(q)}\psi_{\sigma^{(q)}}\right]\otimes \left[\bigotimes_{p'=q+1}^{\NQL}\psi_{\sigma^{(p')}}\right]\nn\\
& = \sum_{q=1}^\NQL \lambda_{\sigma^{(q)}}^{(q)}\psi_{\bs \sigma} = \lambda_{\bs \sigma} \psi_{\bs \sigma},
\end{align}
where 
\begin{equation}\label{eq:lambda_sigma}
\lambda_{\bs\sigma} = \sum_{q=1}^\NQL \lambda_{\sigma^{(q)}}^{(q)}.
\end{equation}
\Cref{eq:R_psi_sigma,eq:lambda_sigma} tell us that the spectrum of the resource $\mc R$ of $\NQL$ QL bits will involve peaks corresponding to sums of the emergent eigenvalues of each $\mc R^{(q)}$. \cite{product}

The states in \cref{eq:psi_sigma} define an orthonormal basis for $\mc H$ according to the scalar product defined by
\begin{align}
\braket{\psi_{\bs \sigma},\psi_{\bs \sigma'}}
&=  \frac 1{2^\NQL}\prod_{q=1}^{\NQL}\left[|a_1^{(q)}|^2+(2\delta_{\sigma^{(q)},\sigma'^{(q)}} -1)|a_2^{(q)}|^2\right]\nn\\ 
&=\prod_{q=1}^{\NQL}\delta_{{\sigma}^{(q)},{\sigma'}^{(q)} }=\prod_{q=1}^{\NQL} \braket{\sigma^{(q)}| {\sigma'}^{(q)}}.
\end{align}
Elementary operators in our approach are defined by 
\begin{align}\label{eq:Psi_ss'}
\Psi_{\bs \sigma,\bs{\sigma}'} &= \psi_{\bs\sigma} \psi_{\bs\sigma'}^\dagger 
\mapsto\bigotimes_{q=1}^{\NQL}  \ket {{\sigma}_q^{(q)}}\bra {{\sigma'}_{q}^{(q)}} = \ket {\bm \sigma}\bra{\bm \sigma'},
\end{align}
The Hermitian conjugates of \cref{eq:Psi_ss'} are defined such that
\begin{equation}\label{eq:Psi_dagger}
\left[\Psi_{\bs \sigma,\bs \sigma'}^\dagger\right]_{\bs{ij}} = \prod_{q=1}^\NQL \psi_{\sigma^{(q)},j^{(q)}}^*\psi_{\sigma'^{(q)},i^{(q)}} = \left[\Psi_{\bs \sigma',\bs \sigma}\right]_{\bs{ij}},
\end{equation}
and they obey the orthogonality relations
\begin{equation}\label{eq:comb_mapN}
\Psi_{\bs{\sigma\sigma'}}\Psi_{\bs{\pi\pi'}} = \Psi_{\bs{\sigma\pi'}}\braket{\bs{\psi_{\sigma'}},\bs{\psi_\pi}}.
\end{equation}

Let us note that the Cartesian product of $\NQL$ QL-bits, each defined by a graph with $2\Ng$ vertices, yields a graph whose adjacency matrix has dimension $(2\Ng)^\NQL$, scaling exponentially with the number of QL-bits. This behavior is, to some extent, an expected trade-off when encoding quantum states into classical many-body systems. As discussed in \cref{sec:perspectives}, this issue presents a practical—though not fundamental—limitation to testing the formalism with available computational and experimental resources.

The mapping proposed in this section provides a general description of state vectors, entanglement and operators constructed from the emergent states of an arbitrary number of QL bits.
In the next \cref{subsubsec:bell} we show explicitly, as an example, how to construct maximally entangled Bell states for $\NQL=2$.

\subsubsection{Example: Bell states}\label{subsubsec:bell}

\begin{figure}
	\centering\includegraphics[width=3.7in]{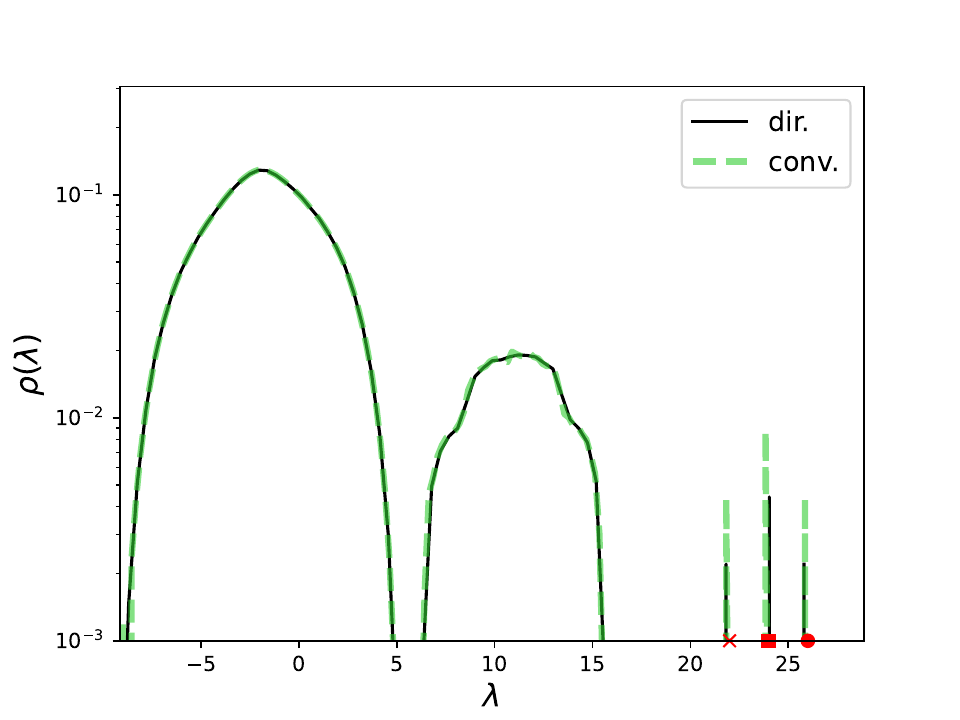}\caption{Distribution of the eigenvalues of \cref{eq:R2}, for  $N_{\mr{samp}} = 10^3$ random samples of the graphs resources. The parameters of the graph are fixed here to $\Ng=16$, $k=6$, $l=1$. The solid black curve corresponds to the direct sampling of the eigenvalues of \cref{eq:R2}, while the green dashed curve is obtained from the convolution of two identical one-QL-bit distributions [\cref{eq:conv}]. }\label{fig:histo_QLtot}
\end{figure}

Here, we study the spectral properties of two-QL-bit resources, and we show how to construct entangled Bell states with them. 
\Cref{eq:R_NQL} in this case simplifies to
\begin{equation}\label{eq:R2}
\mc R =\mc  R^{(1)}\otimes \mathbb 1_{2\Ng} +\mathbb 1_{2\Ng}\otimes\mc  R^{(2)}.
\end{equation}
The spectrum $\rho(\lambda)$ of a resource with structure as in \cref{eq:R2} is shown in \cref{fig:histo_QLtot} for the system parameters $\Ng=16$, $k=6$ and $l=1$.  
To improve the smoothness of all distributions, we average over a number of $N_{\mr{samp}} = 10^3$ realizations of random regular graphs.
The black curve corresponds to the direct (dir.) calculation from the eigenvalues of \cref{eq:R2}.
An alternative approach to calculate the spectrum is possible, by observing that each eigenvalue $\lambda_i$ of \cref{eq:R2} can be expressed as a sum $\lambda_i= \lambda^{(1)}_{j}+ \lambda^{(2)}_{k}$, where the $\lambda^{(q)}_m$'s are eigenvalues of $\mc R^{(q)}$, $q=1,2$.\cite{product}
We can therefore calculate the spectrum of \cref{eq:R2} via the convolution
\begin{equation}\label{eq:conv}
\rho(\lambda) = \int_{-\infty}^{+\infty} \mr d x \;  \rho^{(1)}(x) \rho^{(2)}(\lambda-x),
\end{equation}
where $  \rho^{(q)}(x)$ is the spectrum of a single QL bit, shown in \cref{fig:histo_QL1}.
The generalization of \cref{eq:conv} provides a simple recipe to construct the spectrum of a system of an arbitrary number of QL bits starting from the spectra for $\NQL=1$. 
The distribution calculated from the convolution (conv.) \cref{eq:conv}, is shown as a green (light) dashed curve in \cref{fig:histo_QLtot}, and it agrees satisfactorily with the direct calculation.
We mark in red on the picture the isolated peaks corresponding to the emergent eigenvalues of the two-QL-bit systems. 
These are $\lambda_{11} = 2k+2l$ (circle), $\lambda_{10} = \lambda_{01}=2k$ (cross) and $\lambda_{00}=2k-2l$ (square). 
These eigenvalues are associated to the corresponding eigenvectors $\{\psi_{\sigma\sigma'}\}_{\sigma,\sigma'\in \mc S}$, defining an orthonormal basis for the Hilbert space of two QL bits. 
To obtain the explicit expression of the eigenvectors, it is convenient to label the eigenvectors of each QL bit as 
\begin{equation}\label{eq:psi_12}
\psi^{(1)}_{\ua/\da} = \begin{pmatrix}
a_1 \\
\pm a_2
\end{pmatrix},\hspace{5mm}
\psi^{(2)}_{\ua/\da} = \begin{pmatrix}
b_1 \\
\pm b_2
\end{pmatrix},
\end{equation}
where $a_i$'s and $b_i$'s are defined as in \cref{eq:psi_pm} for the resources $ {\mc R}^{(1)}$ and $ {\mc R}^{(2)}$, respectively.
The notation in \cref{eq:psi_12} indicate that the arrow on the left (right) of the slash is associated to the upper (lower) sign on the right-hands side.
With \cref{eq:psi_12} we can express
\begin{subequations}\label{eq:fact2} 
	\begin{align}
		\psi_{\ua,\ua/\da} &= \frac 12
		\begin{pmatrix}
			a_1\otimes b_1 \\\pm a_1 \otimes b_2\\
			a_2\otimes b_1 \\ \pm a_2 \otimes b_2
		\end{pmatrix},\hspace{5mm}
		\psi_{\da,\ua/\da}&=\frac 12
		\begin{pmatrix}
			a_1\otimes b_1 \\ \pm a_1 \otimes b_2\\
			-a_2\otimes b_1 \\ \mp a_2 \otimes b_2
		\end{pmatrix}.
	\end{align}
\end{subequations}
We can finally construct maximally entangled Bell states via the linear combinations
\begin{subequations}\label{eq:bell}
\begin{align}
\Phi^{\mr{Bell}}_{\ua/\da} &=\tfrac 1 {\sqrt 2}\left(\psi_{\ua\ua} \pm \psi_{\da\da}\right),
 \label{eq:phibell_pm}\\
\Psi^{\mr{Bell}}_{\ua/\da}&=\tfrac 1 {\sqrt 2}\left(\psi_{\ua\da} \pm \psi_{\da\ua}\right), \label{eq:phibell_pm}
\end{align}
\end{subequations}
that is,
\begin{subequations}
	\begin{align}
		\Phi^{\mr{Bell}}_{\ua} &= \frac{1}{\sqrt{2}} \begin{pmatrix}
			a_1 \otimes b_1 \\ 0_{N_{\mr G}^2} \\
			0_{N_{\mr G}^2} \\ a_2 \otimes b_2
		\end{pmatrix}, &
		\Phi^{\mr{Bell}}_{\da} &= \frac{1}{\sqrt{2}} \begin{pmatrix}
			0_{N_{\mr G}^2} \\ a_1 \otimes b_2 \\
			a_2 \otimes b_1 \\ 0_{N_{\mr G}^2}
		\end{pmatrix}, \\
		\Psi^{\mr{Bell}}_{\ua} &= \frac{1}{\sqrt{2}} \begin{pmatrix}
			a_1 \otimes b_1 \\ 0_{N_{\mr G}^2} \\
			0_{N_{\mr G}^2} \\ -a_2 \otimes b_2
		\end{pmatrix}, &
		\Psi^{\mr{Bell}}_{\da} &= \frac{1}{\sqrt{2}} \begin{pmatrix}
			0_{N_{\mr G}^2} \\ -a_1 \otimes b_2 \\
			a_2 \otimes b_1 \\ 0_{N_{\mr G}^2}
		\end{pmatrix},
	\end{align}
\end{subequations}
where
\begin{equation}\label{eq:xM}
	x_M = (x,\cdots, x)/\sqrt{M} \in \mathbb C^{M}, \hspace{10mm}x \in \mathbb C,\; M \in \mathbb N.
\end{equation}

So far, we have presented general results on the QL formalism and demonstrated that the approach enables an exact mapping to the Hilbert space of a qubit system.
In the following \cref{sec:sync_dyn}, we explore how the abstract concepts of graph resources and emergent states relate to the classical dynamics of a system of oscillators. We discuss the resilience of this approach in the presence of noise in the network correlations and highlight the critical role of emergent states in the classical synchronizing dynamics.

\section{Classical synchronization}\label{sec:sync_dyn}

\begin{figure*}
\includegraphics[width=1.\linewidth]{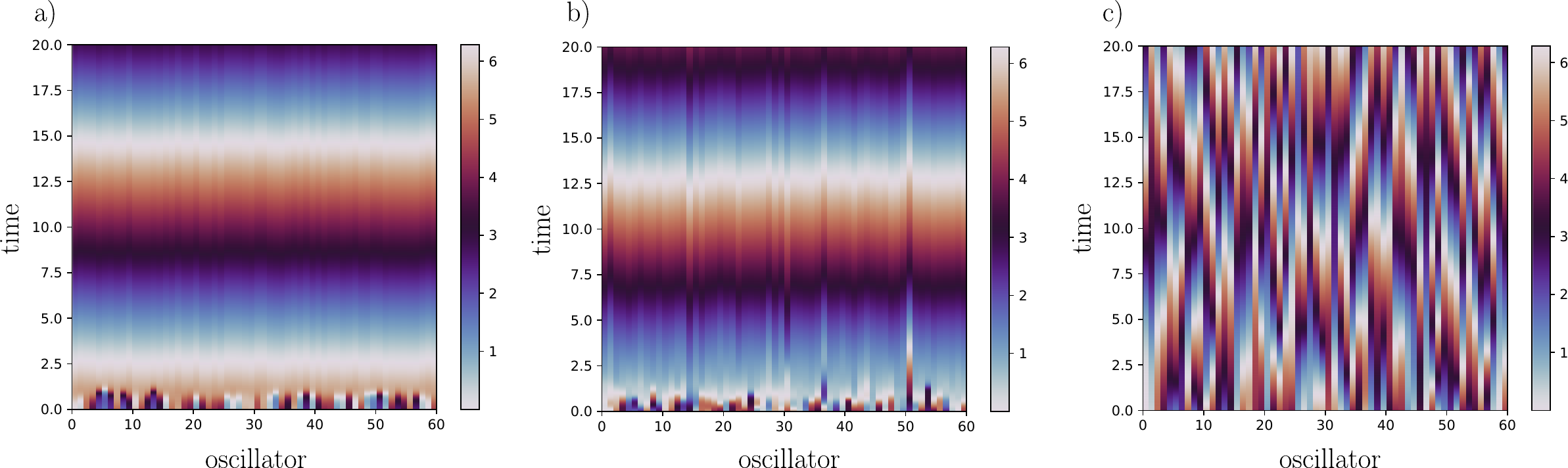}
\caption{
a): Time evolution of a system with $2\Ng = 60$ oscillators governed by the Kuramoto model [\cref{eq:kuramoto}]. The angular variables $\theta_i$ are initialized uniformly in $[0, 2\pi)$, and the intrinsic frequencies $\omega_i$ are sampled uniformly from $[0, 1]$. The coupling constant is set to $\Gamma = 10$, and the connectivity matrix $\mc{R}$ is structured as a regular graph, as described in \cref{eq:R}, with the valencies of the diagonal and off-diagonal blocks set to $k=29$ and $l=15$, respectively.
b): Time evolution of the system under the same parameters and initial conditions as in a), but with disorder introduced into the connectivity matrix $\mc{R}$. Specifically, each nonzero entry of $\mc{R}$ is perturbed by random white noise with zero mean and variance $\sigma_{\mr n} = 3$.
c): Dynamics of the Kuramoto model with the connectivity matrix restricted to its random components [see discussion following \cref{eq:R_em_ran}]. The system parameters and initial conditions are the same as in a) and b).}
\label{fig:kuramoto1}
\end{figure*}

Here, we investigate classical synchronization in dynamical systems of oscillators, where two-body correlations are governed by a connectivity matrix with the structure described in \cref{eq:R}.
Systems that align with our formalism must, as a general requirement, involve classical many-body dynamics, where the two-body couplings between degrees of freedom are governed by a connectivity matrix $\mc R$, as outlined in \cref{sec:QLbits}. Additionally, the dynamics should guarantee that synchronization leads to a globally stable state that remains robust over long time scales.
These features are satisfied by the Kuramoto model, \cite{kuramoto1975,kawamura2010,scholes2022,rodrigues2016} described by the nonlinear and non-symplectic equations of motion
\begin{equation}\label{eq:kuramoto}
	\dot \theta_i = \omega_i + \frac{\Gamma}{2\Ng}\sum_{j=1}^{2\Ng} \mc R_{ij}\sin(\theta_j-\theta_i), \hspace{7mm} i=1,\ldots,2\Ng.
\end{equation}
Here, the $\theta_i$'s are angular variables corresponding to $2\Ng$ oscillators, defined on the vertices of a graph. The matrix $\mc R$ encodes the correlations between these degrees of freedom. 
The $\omega_i$'s denote the intrinsic oscillation frequencies.
The time evolution of the angular variables of the model is shown in panel a) of \cref{fig:kuramoto1}.
For our chosen system parameters (see caption of the figure) the system reaches a synchronized steady state,  where all oscillators evolve at a uniform, constant phase.
The synchronized steady state appears robust even in presence of disorder in the network, as shown in panel b) of the same figure. 
Here, white static disorder, with variance $\sigma = 3$ and a zero mean, is added to all nonzero entries of the adjacency matrix $\mc R$, originally valued at one. 
Despite this strong perturbation, a collective synchronization state persists at long times, with only minor phase shifts affecting small subgroups of oscillators.
This observation further supports the authors' previous findings on the resilience of synchronization dynamics to noise---a property particularly appealing for robust  information encoding and processing in the steady states of these classical dynamical systems. \cite{product,scholes2023,scholes2024}

To examine the role of the emergent states in the synchronization dynamics it is convenient to express the adjacency matrix through its spectral decomposition, as
\begin{equation}\label{eq:R_em_ran}
\mc R = \sum_{\bs \sigma \in \bs{\mc S}} \lambda_{\bs\sigma} \psi_{\bs\sigma}  \psi_{\bs \sigma}^\dagger + \sum_{i=1}^M \tilde \lambda_{i}\tilde\psi_i  \tilde\psi_i^\dagger,
\end{equation}
where $M=(2\Ng)^\NQL-2^\NQL$.
The first term in \cref{eq:R_em_ran} represents the contribution of the emergent states $\psi_{\bs \sigma}$'s.  
The second term corresponds to the random part of the spectrum, with eigenvalues $\tilde \lambda_i$'s and corresponding eigenvectors $\tilde \psi_i$'s.
In panel b) of \cref{fig:kuramoto1}, we show the time evolution of a system initialized with the same conditions as in panel a) and governed by the same interaction rules, with the only difference being the addition of static random noise \( n \). This noise has zero mean, variance \( \sigma_{\mathrm{n}} = 3 \), and is sampled independently for each nonzero component of the adjacency matrix.
Despite the noise being strong compared to the oscillator coupling strength (set to 1), the system still shows a clear synchronization pattern over time. 
This provides numerical evidence that, under broad conditions, the dynamics of the model can be highly resilient to static disorder. Such robustness is related to the collective behavior that emerges through synchronization.
Finally, we show in panel c) of \cref{fig:kuramoto1} the time evolution of the Kuramoto dynamics when only the random part of the spectrum is taken into account [that is, by imposing $\lambda_{\bs \sigma}\equiv 0$ $\forall \bs \sigma \in \bs{\mc S}$ in \cref{eq:R_em_ran}].
In this scenario, synchronization fails to emerge, with all degrees of freedom maintaining their initial random phases throughout the entire time evolution. 
This highlights that the structure of two-body correlations and emergent states, encoded in the adjacency matrix $\mc R$, play a critical role in achieving classical synchronization.
In the next \cref{sec:gates}, we build on these idea by exploring the connection between dynamical synchronization and information processing. 
This approach provides a framework for linking the steady states of classical many-body systems to quantum information protocols.

\section{Quantum-like gates}\label{sec:gates}

In this section, we explore the fundamental principles of QL information processing.
In \cref{subsec:gates_f}, we formalize how the emergent states of QL resources can be systematically transformed according to arbitrary quantum gates. 
We further investigate the correspondence between state transformations and network resources.
In \cref{subsec:gates_s}, we connect this abstract formalism to the dynamics of classical synchronization.
Specifically, we demonstrate how QL gates can be practically realized in classical many-body systems of synchronizing oscillators and analyze the impact of these transformations on the long-time steady states of their dynamics.

\subsection{Formalism}\label{subsec:gates_f}

To connect the usual definition of gates acting on an abstract Hilbert space to our QL formalism, it is convenient to introduce a map $U_{\mr{cb}}$ transforming the emergent eigenvectors onto the standard  computational basis.
Given that all elementary subrgaphs in our resources [\cref{eq:R,eq:R_NQL}] are regular, this map can be expressed analytically by
\begin{equation}\label{eq:Ucb}
		U_{\mr{cb}} =  [V_{\mr H}\otimes \mathbb 1_\Ng]^{\otimes \NQL},
	\end{equation}
	where
	\begin{equation}
		V_{\mr H} = \frac 1 {\sqrt 2}\begin{pmatrix}
			1 & 1 \\
			1 & -1
		\end{pmatrix}
	\end{equation}
	denotes the matrix representation of the Hadamard (H) gate in the computational basis.
The action of $U_{\mr{cb}}$ becomes clear for $\NQL = 1$, where
\begin{equation}\label{eq:Ucb_1}
U_{\mr{cb}}\psi_\ua = \begin{pmatrix}  1_\Ng \\  0_\Ng \end{pmatrix}, \hspace{5mm} U_{\mr{cb}}\psi_\da = \begin{pmatrix}  0_\Ng \\  1_\Ng \end{pmatrix}.
\end{equation}
Note that, given that all the basis subgraphs are regular, the solution to \cref{eq:Ucb_1} is independent of their local structure. We show in \cref{app:Ucb} that a solution for the map $U_{\mr{cb}}$ can be determined even if the resources are not exactly regular, by solving numerically a linear algebra problem.
We define the QL analog of any arbitrary quantum gate $V_{\mr g}$ by the matrix transformation 
\begin{equation}\label{eq:Ug}
	U_{\mr g} =  U_{\mr{cb}}^{-1} (V_{\mr g} \otimes  \mathbb 1_{\Ng}) U_{\mr{cb}}.	
\end{equation}
To elucidate the action of \cref{eq:Ug}, let us consider a few examples. 
The QL maps of the identity operator ($\mc I$) and the three Pauli gates (X,Y,Z) are
\begin{subequations}
\begin{align}
U_{\mc I}&= 
\begin{pmatrix}
	\mathbb 1_\Ng & \mathbb 0_\Ng \\
	\mathbb 0_\Ng & \mathbb 1_\Ng
\end{pmatrix}, \label{eq:UI}\\
U_{\mr X} &= 
\begin{pmatrix}
\mathbb 1_\Ng & \mathbb 0_\Ng \\
 \mathbb 0_\Ng & -\mathbb 1_\Ng
\end{pmatrix}, \label{eq:UX}\\
U_{\mr Y} &=
\begin{pmatrix}
	\mathbb 0_\Ng & i\mathbb 1_\Ng \\
	-i\mathbb 1_\Ng & \mathbb 0_\Ng
\end{pmatrix}, \label{eq:UY}\\
U_{\mr Z} &= \begin{pmatrix}
	\mathbb 0_\Ng & \mathbb 1_\Ng 	\\
	\mathbb 1_\Ng & \mathbb 0_\Ng 
\end{pmatrix},
\end{align}
\end{subequations}
respectively. As expected, the transformations in \cref{eq:UX} act as the standard Pauli gates, namely,
\begin{equation}\label{eq:act_pauli}
U_{\mr X} \psi_{\ua/\da} = \psi_{\da/\ua}, \hspace{5mm} U_{\mr Y} \psi_{\ua/\da} = \pm i \psi_{\da/\ua}, \hspace{5mm} U_{\mr Z} \psi_{\ua/\da} = \pm \psi_{\ua/\da}.
\end{equation}
As another example, the QL map corresponding to the Hadamard gate transforms states in the factorized basis into superposition states:
\begin{equation}\label{eq:UH}
	U_{\mr H} \psi_{\ua/\da} = \left(V_{\mr H}\otimes \mathbb 1_{\Ng}\right) \psi_{\ua/\da} = \frac 1 {\sqrt 2}(\psi_\ua\pm \psi_\da).
\end{equation} 
The approach proposed here applies equally to any kind of gate, including those outside the Clifford group.
This class of gates break the stabilizer structure, making them harder to simulate compared to Clifford gates.
In contrast, Clifford gates can be simulated efficiently on a classical computer.
However, this difference between Clifford and non-Clifford gates does not matter in our framework.
Our formalism already takes into account the exponential cost of scaling QL resources, no matter what type of gate is used.
The action of the non-Clifford T-gate on the factorized basis is
\begin{equation}
U_{\mr T}\psi_{\sigma} = \left(\delta_{\sigma,\ua} + \e^{i\pi/4} \delta_{\sigma,\da}\right)\psi_\sigma.
\end{equation}
Finally, as an example of two-QL-bit transformation, we consider the controlled NOT (CN) gate. This is defined in an abstract Hilbert space by $ \hat V_{\mr {CN}} = \ket \ua \bra \ua \otimes \hat {\mc I} + \ket \da \bra \da \otimes \hat{\mr X}$, where $\hat {\mc I}$ and $\hat {\mr X}$ are the identity and the Pauli-X operator, respectively. 
As expected, the transformation acts on the second QL bit of a two-QL-bit Hilbert space as either the identity or the NOT gate, depending on the control state defined by the first QL bit. The corresponding QL operator
\begin{align}
U_{\mr {CN}} &= \frac 12 (U_{\mc I}+U_{\mr Z}) \otimes U_{\mc I} + \frac 12 (U_{\mc I}-U_{\mr Z}) \otimes U_{\mr X}
\end{align}
transforms the elements of the factorized basis as follows:
\begin{equation}\label{eq:CNOT}
	U_{\mr{CN}} \psi_{\ua,\ua/\da} = 	\psi_{\ua,\ua/\da},\hspace{10mm}U_{\mr{CN}} \psi_{\da,\ua/\da} = 	\psi_{\da,\da/\ua}.
\end{equation}

Up to this point, we have focused exclusively on the action of QL gates on states $\psi_{\bs\sigma}$ of the factorized basis. 
We can infer from those the transformation of any arbitrary state $\psi = \sum_{\bs\sigma\in\bs{\mc S}} \gamma_{\bs\sigma}\psi_{\bs\sigma}$, $\gamma_{\bs\sigma} \in \mathbb C$ from the linearity of the map \cref{eq:Ug}.
From a practical standpoint, it is  preferable to translate these state operations onto transformations applicable directed to the graph resources. This feature is a key requirement for a future experimental validation of our formalism with experimental resources. (An extended discussion on experimental resources is provided in the next \cref{sec:perspectives}).
A straightforward approach in this regard is to define the unitary map 
\begin{equation}\label{eq:R_to_Rg}
	\mc R \mapsto \mc R_{\mr g} = U_{\mr g} \mc R  U_{\mr g}^\dagger.
\end{equation}
\Cref{eq:R_to_Rg} transforms all eigenstates of a resource consistently to the gate $U_{\mr g}$ as defined in \cref{eq:Ug}.
This can be easily inferred by expanding $\mc R$ on the right-hand side of \cref{eq:R_to_Rg} via the spectral decomposition in \cref{eq:R_em_ran}. 
This map can be straightforwardly extended to arbitrary quantum circuits, described by the sequence of $M$ gate operations $\bs {\mr g} = \{\mr g_1,\ldots,\mr g_M\}$.
In this case, we define by 
\begin{equation}\label{eq:Ug}
U_{\bs {\mr g}} =  U_{\mr{cb}}^{-1} (V_{\bs {\mr g}} \otimes  \mathbb 1_{\Ng}) U_{\mr{cb}} = U_{\bs {\mr g}}= U_{\mr g_M}\cdots U_{\mr g_1},
\end{equation}
the corresponding unitary map acting on QL states, with an analog network transformation defined by 
\begin{equation}\label{eq:Rg_circ}
\mc R_{\bs {\mr g}} = U_{\bs {\mr g}} \mc R U_{\bs {\mr g}}^\dagger.
\end{equation}

In the next \cref{subsec:gates_s} we provide concrete examples of \cref{eq:R_to_Rg}, and we discuss how this network transformation affects the steady-state synchronizing solution of a system of oscillators.
A more general analysis of circuit transformations, associated to \cref{eq:Rg_circ}, can be found in Ref.~\onlinecite{encoding}.

\subsection{Connection to classical synchronization}\label{subsec:gates_s}

Here, we explore the relationship between gate transformations of QL resources and the synchronization of classical many-body oscillator systems. Specifically, we focus on two examples of one-QL-bit gates. 
In these cases, the synchronization patterns of the classical system can be directly linked to the structure of the corresponding adjacency matrices. However, our approach is general and can be extended to arbitrary QL gates without any restrictions.

\begin{figure}
\includegraphics[width=0.8\linewidth]{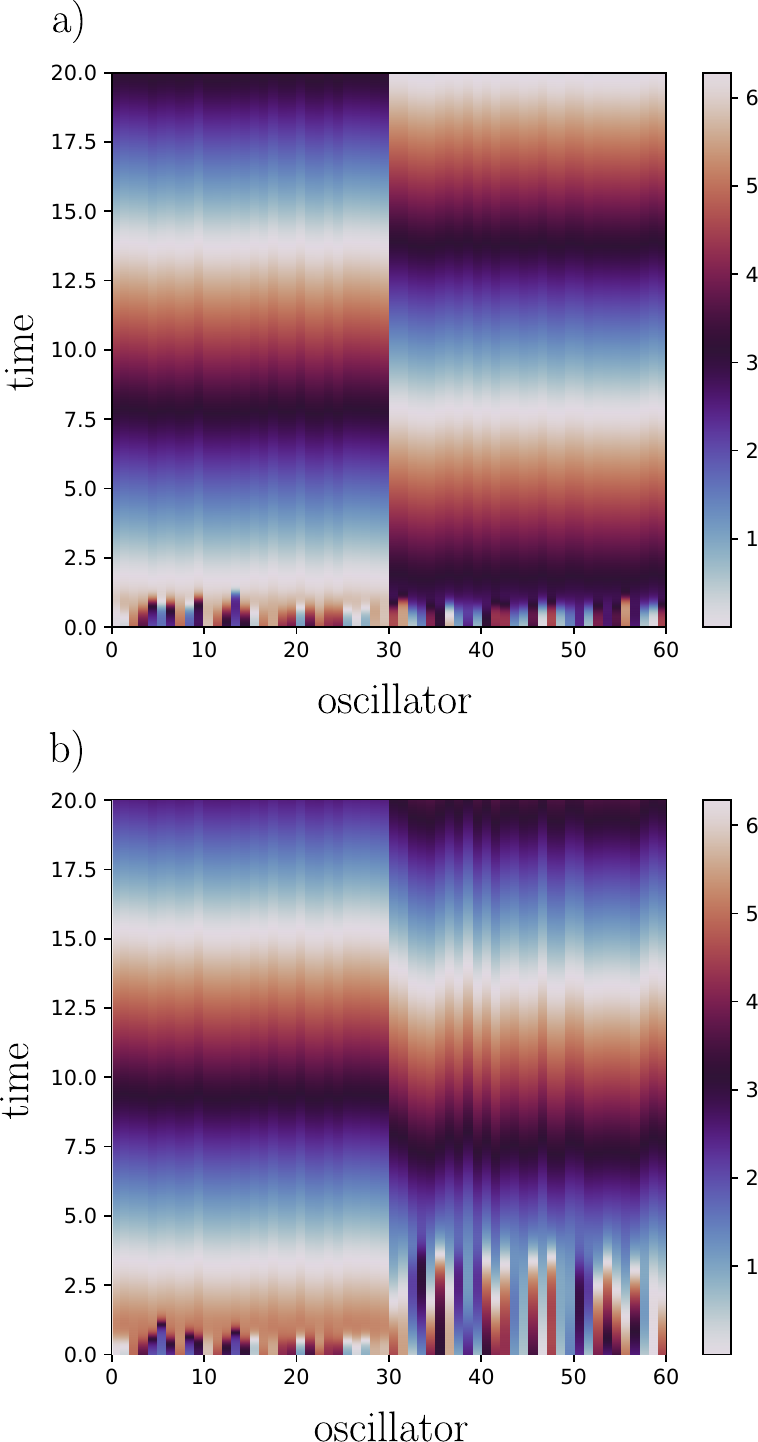}
\caption{
a): Time evolution of a Kuramoto model where the adjacency matrix is transformed according to the mapping in \cref{eq:R_to_Rg} for the Pauli-X gate (g = X).
b): Similar to panel (a), but for the Hadamard gate (g = H).
The same model parameters as defined in \cref{fig:kuramoto1} are used here.
}\label{fig:kuramoto2}
\end{figure}

The Pauli-X (NOT) gate transforms an initial resource \cref{eq:R} as follows:
\begin{equation}\label{eq:RX}
\mc R \mapsto \mc R_{\mr X} 
= \begin{pmatrix} A_1 & -C \\ -C^\top & A_2\end{pmatrix}.
\end{equation} 
We show in panel a) of \cref{fig:kuramoto2} the synchronization pattern in the dynamics of the Kuramoto model with correlation matrix $\mc R_{\mr X}$.
In the transformation \cref{eq:RX}, the leading emergent eigenvector, associated to the largest eigenvalue $k+l$, is mapped from $\psi_\ua $ to $\psi_\da$ [see \cref{eq:psi_pm}].
The change of sign in the eigenvectors corresponds to a phase shift of $\Delta \phi=\pi$ between the first and the second half of the oscillators, as shown in the figure.
In the case of the Hadamard gate, resources are transformed according to
\begin{equation}\label{eq:RH}
\mc R \mapsto \mc R_{\mr H} = \begin{pmatrix} A_1+A_2+C+C^\top & A_1-A_2-C+C^\top \\ A_1-A_2+C-C^\top & A_1+A_2-C-C^\top \end{pmatrix}.
\end{equation}
The structure of \cref{eq:RH} encodes information about the expected synchronization patterns in the dynamics of the corresponding Kuramoto model, as depicted in panel b) of \cref{fig:kuramoto2}. 
Dephasing emerges between the first and second halves of the oscillators, primarily driven by cancellation effects in the two off-diagonal blocks.
When $A_1 \approx A_2$ and $C \approx C^\top$, these blocks approach zero, decorrelating the phases between the two oscillator subgroups.
Furthermore, the second half of the oscillators, labeled by indices $i\in\{\Ng,\ldots, 2\Ng\}$ exhibits weaker synchronization compared to the first half. 
This is due to the fact that the matrix norm of the lower diagonal block in \cref{eq:RH} is, on average, smaller than that of the upper diagonal block. Consequently, the second group of oscillators exhibits weaker correlations compared to the first group.
The interested reader is referred to Ref.~\onlinecite{encoding}. There, we show with a straightforward generalization of the present Kuramoto model  that this kind classical synchronization patterns can accurately represent arbitrary quantum states.

In this section, we derived general principles for mapping network resources in a manner consistent with an arbitrary set of QL gates. 
Additionally, we presented numerical evidence supporting the connection between QL gates and the steady-state synchronization of classical oscillator resources.
In the following \cref{sec:perspectives}, we explore specific examples of experimental resources that could be utilized to implement this approach in practice. 

\section{Experimental perspectives}\label{sec:perspectives}

So far, we have shown that the emergent states of synchronizing classical systems can be used as a basis to map the Hilbert space of a system of qubits. 
We discussed the primary role of emergent states in synchronization dynamics and the robustness of the latter in the presence of noise.
Additionally, we designed a general approach to implement gate transformations on the network resources.
In this section, we explore perspectives on implementing the proposed formalism using experimental physical resources. 
Additionally, we provide a high-level outline of computational and experimental validations that could support the further development of this framework.

For genuine quantum resources, the primary source of errors during information processing and measurement arises from uncontrolled noise. 
This noise can result from several factors, including external electromagnetic fields, unwanted interactions between qubits, fabrication defects, and imperfections in control protocols. \cite{hou2016, bengtsson2024, funcke2022, gupta2021, salonik2021} 
Addressing such errors typically requires multiple measurements to accurately resolve the final state of a quantum register after computations.
This can introduce significant overhead in computing protocols and hinder the scalability of the approach to larger size. 
Error mitigation and decoding techniques, although conceptually scalable, can result in further limitations in size and performance. \cite{bravyi2021, ali2024}
In our approach, error correction in the presence of noise could be achieved by encoding information into a larger number of oscillators. Given that those are defined on a classical phase space, the related cost of the strategy scales polynomially with the number of resources. This suggests a simple strategy to implement a QL version of repetition codes in our approach.

In terms of resource complexity, the number of degrees of freedom required to model the synchronization dynamics of a system of $\NQL$ QL bits scales exponentially with system size. 
Although this scaling does not impose fundamental constraints on computational studies or experimental implementations of the approach, it does limit the practical feasibility of large-scale QL algorithms.
Here, we discuss a few experimental platforms that are suitable candidates for implementing QL information processing within these scaling constraints. 
Additionally, we identify the classes of algorithms that can be most effectively analyzed by those resources.

Implementing coupled LC circuits is a straightforward approach to build macroscopic resources that exhibit classical synchronization. \cite{allam2006,barbi2021}
Numerical analyses from Ref.~\onlinecite{maffezzoni2015} show that this class of systems can be effectively approximated by Kuramoto-type models for medium-scale systems of 60 oscillators.
The couplings between the model oscillators are tuned by adjusting the transconductance  of the differential pair transistors.

Spin-torque oscillators (STO) emerge as another promising candidate for implementing QL information processing. \cite{raychowdhury2019}
STOs are nonlinear magnetic oscillators that can be integrated into nanoscale devices operating at microwave frequencies. 
They have the ability to synchronize in both frequency and phase with external oscillatory signals as well as with other STOs.
The full magnetization dynamics of these systems, defined by the Landau-Lifshitz-Gilbert-Slonzewski equation, can be simplified to the Kuramoto model for small-amplitude oscillations. \cite{belanovsky2013,kanao2019}
Ref.~\onlinecite{flovik2016} demonstrates the experimental feasibility of modeling synchronizing classical systems with up to a few hundred STOs. 
Beyond this scale, technical constraints arising from finite-size effects restrict the extent of achievable synchronization.

Metal-to-insulator transition (MIT) devices leverage materials that can switch between insulating and metallic (conductive) states in response to external conditions such as temperature, pressure, or electrical stimuli. \cite{kim2009}
Research on small-scale MIT circuits has demonstrated their fundamental ability to produce classical synchronization behavior through voltage-triggered oscillatory dynamics. \cite{kim2009,parihar2015,raychowdhury2019}
One limitation of current MIT devices is their low oscillation frequency, typically below 1 MHz, which hinders the ability to achieve fine-resolution synchronization and detect subtle variations in coupled oscillator behavior.  
However, ongoing advancements in scaling down these devices are expected to increase their oscillation frequencies, enabling more accurate time-resolved synchronization measurements. 
\cite{cotter2014}

Building on the perspectives discussed above, experimental validation of our theory using networks of up to a hundred oscillators appears feasible. In such a setup, each QL bit could consist of a small network of 5–10 oscillators, allowing for proof-of-concept implementations with $\NQL=2$ or $\NQL=3$ QL bits. At this scale, a particularly interesting class of quantum information processes includes entanglement generation, such as the formation of Bell states (for $\NQL=2$) and their higher-order generalizations, the Greenberger–Horne–Zeilinger (GHZ) states. \cite{greenberger1989,erhard2020}
Advancements in this framework are given in Ref.~\onlinecite{encoding}.

Exploring small-scale quantum processes might offer a possible route to address some foundational questions. For example: To what extent can one construct a classical analogue of entanglement in a classical many-body system? Is there a fundamental limit beyond which even infinite classical resources fail to replicate essential features of entanglement? Such a limitation could signal an intrinsic boundary between which kind of quantum behaviors can be reproduced by classical dynamics. We argue that our approach can serve as a platform to investigate these general problems.
\\

We end this section with a practical outlook on how our approach could be tested experimentally.
\begin{enumerate} 
\item \emph{Engineer a classical synchronizing system:} Design a classical system of nonlinear oscillators with two-body interactions, defined by a graph. This graph should be constructed so that its leading emergent state matches the desired initial state of a quantum circuit. For example, the factorized state $\psi_\ua$ corresponds to the eigenvector with the largest eigenvalue of a Cartesian-product network, as given in \cref{eq:R_NQL}. The other parameters of the classical system should be tuned to ensure the emergence of synchronization at long times.
\item \emph{Transform the network according to arbitrary quantum logic}: Apply to the network a unitary transformation corresponding to a given quantum circuit, following the protocol proposed in \cref{subsec:gates_f}.
\item \emph{Allow the classical dynamics to equilibrate:} Assign random initial phases to the classical oscillators and let the system evolve until it reaches a synchronized steady state. We discuss formally in Ref.~\onlinecite{encoding} how this steady states encodes information on the gate initially applied to the classical network.
\item  \emph{Measure the final state:} 
Calculate the overlap between the oscillators' steady state and the eigenstates of arbitrary observable operators.
The resulting projection coefficients can then be interpreted as nonclassical probability amplitudes within the framework of QL theories, as discussed in detail in the next \cref{sec:quantum_prob}.
\end{enumerate}
In the next and final \cref{sec:quantum_prob} we expand over this last point, to show that our approach allows to mimic non-Kolmogorov statistics, that violates the Bayes theorem due to the occurrence of nonclassical interference.


\section{Quantum-like interfence}\label{sec:quantum_prob}

In this final section, we discuss criteria that allow to introduce a notion of QL interference of probabilities in our approach.
This analysis highlights that genuine quantum mechanical systems are not strictly required to build QL stochastic models provided the notions of states, observables and measurement are defined according to a given set of prescriptions.

\subsection{Axiomatic definition}\label{subsec:axioms}

In the following we present a list of axioms, originally proposed in Ref.~\onlinecite{khrennikov2010}, to equip our classical network resources with a stochastic theory which mirrors the probability laws of genuine quantum mechanical systems.
 
\begin{axiom}\label{axiom:1}
QL states are represented by normalized vectors in a Hilbert space $\mc H$.
\end{axiom}
States can be expressed by linear combinations of elements of the factorized basis in \cref{eq:psi_sigma}:
\begin{equation}\label{eq:psi_pure}
\psi = \sum_{\bs\sigma\in\bs {\mc S} }\alpha_{\bs \sigma} \psi_{\bs\sigma}.
\end{equation}
Pure QL states are by definition normalized, and we establish the equivalence class $\beta \psi \sim \psi$ for $\beta \in \mathbb C$, $|\beta|=1$. 

\begin{axiom}\label{axiom:2}
QL observables are represented by self-adjoint operators on $\mc H$. Their action on QL states, expressed by a linear map, defines a QL measurement.
\end{axiom}
Observables are defined in terms of linear combinations with structure
\begin{align}\label{eq:F}
	\mc {F} &= \sum_{{\bs\sigma,\bs\sigma'\in \bs{\mc S}} }\left(\gamma_{ \bs\sigma, \bs\sigma'}\Psi_{\bs\sigma,\bs\sigma'} +\gamma^*_{\bs\sigma,\bs\sigma'}\Psi_{\bs\sigma',\bs\sigma} \right), \hspace{5mm} \gamma_{\bs\sigma,\bs\sigma'}\in \mathbb C,
\end{align}
and fulfill the self-adjointness condition $\mc F^\dagger=\mc F$ [\cref{eq:Psi_dagger}].
\begin{axiom}\label{axiom:3}
	The spectrum of a QL observable  is a measurable quantity.
\end{axiom}

An arbitrary observable $\mc F$, as defined in \cref{axiom:2}, is diagonalizable.
Its normalized eigenvectors {$\{f_{\bs\sigma}\}$$_{\bs\sigma \in \bs{\mc S}}$} define an orthonormal basis of $\mc H$, while the corresponding eigenvalues are real and distinct. 
Proofs of these standard properties for the specific structure of our QL Hilbert space are given in \cref{app:spectrum}.

\begin{axiom}\label{axiom:4}
A QL probability measure defined by Born's rule is assigned to the spectrum of physical observables.
\end{axiom}
Specifically, let us consider a generic state $\psi\in \mc H$.
We define by 
\begin{equation}\label{eq:born}
P_\psi\left(\lambda_{\bs \sigma}^{\mc F}\right) = \norm{\Pi^{\mc F}_{\bs\sigma} \psi}^2 
= |\braket{f_{\bs \sigma}, \psi}|^2,
\end{equation}
the QL probability of collapsing onto the subspace spanned by the normalized eigenvector $f_{\bs \sigma}$ with eigenvalue $\lambda^{\mc F}_{\bs \sigma}$, after measuring $\mc F$ on $\psi$.
\Cref{eq:born} is expressed in terms of the projector
\begin{equation}\label{eq:Pi_psi}
	\Pi^{\mc F}_{\bs \sigma} = f_{\bs \sigma} f_{\bs \sigma}^\dagger = \left(\Pi^{\mc F}_{\bs \sigma}\right)^2.
\end{equation}

\begin{axiom}\label{axiom:5}
QL conditional probabilities are defined via consecutive projections onto the eigenstates of physical observables.
\end{axiom}
In particular, let us consider two generic QL observable operators $\mc F$ and $\mc G$ whose spectra decompositions are defined by
\begin{equation}\label{eq:eigFG}
\mc F = \sum_{\bs \sigma \in {\bs{\mc S}}}\lambda_{\bs\sigma}^{\mc F}f_{\bs \sigma} f_{\bs\sigma}^\dagger, \hspace{10mm}  
\mc G = \sum_{\bs \sigma \in {\bs{\mc S}}}\lambda_{\bs\sigma}^{\mc G}g_{\bs \sigma} g_{\bs\sigma}^\dagger,
\end{equation}
respectively.
For a nonzero state 
$\psi$ we denote by
\begin{equation}\label{eq:P_psi_cond}
	P_\psi\left(\lambda^{\mc G}_{\bs \sigma'}|\lambda^{\mc F}_{\bs \sigma}\right) = \frac{\norm{\Pi^{\mc G}_{\bs \sigma'} \Pi^{\mc F}_{\bs \sigma}  \psi}^2}{\norm{\Pi^{\mc F}_{\bs \sigma} \psi }^2}
\end{equation}
the QL conditional probabilities associated to the consecutive measures of $\mc F$ and $\mc G$ on  $\psi$.
\\

We show in the next \cref{subsec:q_prob} that the stochastic model defined via \crefrange{axiom:1}{axiom:5} violates by construction Kolmogorovian classical statistics.
This feature allows to introduce a notion of QL interference in our approach.
\cite{khrennikov2003,khrennikov2019}

\subsection{Quantum probabilities in a QL model}\label{subsec:q_prob}

Before introducing a formula for the update of QL probabilities,
it is useful to briefly recall its classical correspondent, to assess analogies and differences between the two. 
A classical probability space is defined by the triplet $\{\Omega, \Sigma, P^{(\mr{cl})}\}$, where $\Omega$ is a space of experimental outcomes, $\Sigma$ the $\sigma$-algebra of its subsets, and $P^{(\mr{cl})}$ a probability measure on $\Sigma$.\cite{shiryaev2013}
Be $\{\mc A_j\}_{j=1}^J$ a partition of $\Omega$, i.e. a collection of $J$ sets defined such that $\mc A_j \neq \emptyset \; \forall j$, $\mc A_j \cap A_{j'} = \emptyset$ for $j\neq j'$ and $\cup_{j=1}^{J}\mc A_j = \Omega$.
$P^{(\mr{cl})}$ is a positive measure on the partition elements, that is $P^{(\mr{cl})} (\mc A_j)>0 \; \forall j$, and it is normalized: $\sum_{j=1}^J  P^{(\mr{cl})} (\mc A_j)=1$.
For every $\mc B\in \Sigma$ the classical formula of total probabilities holds: \cite{kolmogoroff1933} 
\begin{align}\label{eq:CFTP}
P^{(\mr{cl})}(\mc B) &= P^{(\mr{cl})}\left(\mc B\cap \cup_{j=1}^J A_j \right) \nn\\
&= \sum_{j=1}^{J}  P^{(\mr{cl})}\left(\mc B\cap A_j \right) = \sum_{j=1}^{J} P^{(\mr{cl})}(\mc B |\mc A_j)P^{(\mr{cl})}(\mc A_j),
\end{align}
where we introduced the classical conditional probability
\begin{equation}\label{eq:cond_class}
P^{(\mr{cl})}(\mc B |\mc A_j) = \frac{P^{(\mr{cl})}(\mc B \cap \mc A_j)}{P^{(\mr{cl})}(\mc A_j)}.
\end{equation}

We discuss now the correspondent of \cref{eq:CFTP} for the present QL framework.
We restrict the following analysis to couples of non-degenerate observables,  $\mc F$ and $\mc G$,  whose eigenvalue problems are defined as in \cref{eq:eigFG}. 
The eigenvalues of these operators are associated with one-dimensional subspaces, which can be fixed by selecting any eigenvector.
In this case, the QL conditional probabilities defined in \cref{eq:P_psi_cond} lose their dependence on the initial state $\psi$, and can be simply expressed as 
\begin{align}\label{eq:symm_cond}
	P_\psi\left(\lambda^{\mc G}_{\bs \sigma'}|\lambda^{\mc F}_{\bs \sigma}\right) &\mapsto |\braket{ g_{\bs \sigma'},f_{\bs \sigma}}|^2\equiv P\left(\lambda^{\mc G}_{\bs \sigma'}|\lambda^{\mc F}_{\bs \sigma}\right) .
\end{align}
Note that the probabilities in \cref{eq:symm_cond} are symmetric, differently from their classical correspondents in \cref{eq:cond_class}.
Observables $\mc F$ and $\mc G$ obeying this property for all $\bs\sigma,\bs\sigma'\in\bs{\mc S}$ are called \emph{symmetrically conditioned}. \cite{khrennikov2010}
This feature is a signature of \emph{Bohr's principle of contextuality},  \cite{bohr1987,khrennikov2005,khrennikov2022}
expressing the impossibility of strictly separating physical observables and the experimental context under which those are measured.
The role of contextuality in a QL probabilitistic  framework can be highlighted by rewriting total and conditional probabilities [\cref{eq:born,eq:symm_cond}, respectively], on an equal footing:
\begin{equation}\label{eq:cond_cont}
	P\left( \lambda^{\mc G}_{\bs \sigma}|\lambda^{\mc F}_{\bs \sigma'} \right)  =\left.P_{\psi}(\lambda^{\mc G}_{\bs \sigma})\right|_{\psi=f_{\bs\sigma'}}.
\end{equation}
If \cref{eq:symm_cond} is fulfilled, 
\crefrange{axiom:1}{axiom:5} lead to the QL formula for probability updates \cite{khrennikov2007interf,khrennikov2010,bulinksi2004}
\begin{align}\label{eq:QFTP}
P_\psi \left(\lambda_{\bs \sigma}^{\mc G}\right) 
&= \sum_{\bs\sigma'\in \bs{\mc S}}P_\psi \left(\lambda_{\bs \sigma'}^{\mc F}\right)P\left(\lambda^{\mc F}_{\bs \sigma'}|\lambda^{\mc G}_{\bs \sigma}\right)  \nn\\
&\quad + \sum_{\substack{\bs\sigma', \bs\sigma''\in \bs{\mc S} \\\bs\sigma'\neq \bs\sigma''}}
\Lambda^\psi_{\bs\sigma,\bs\sigma',\bs\sigma''}\Big[P_\psi \left(\lambda_{\bs \sigma'}^{\mc F}\right)P\left(\lambda^{\mc F}_{\bs \sigma'}|\lambda^{\mc G}_{\bs \sigma}\right)\nn\\
&\quad\times P_\psi \left(\lambda_{\bs \sigma''}^{\mc F}\right)P\left(\lambda^{\mc F}_{\bs \sigma''}|\lambda^{\mc G}_{\bs \sigma}\right)\Big]^{1/2}.
\end{align}
An extended derivation of \cref{eq:QFTP} is provided in \cref{app:QFTP}, where we give the explicit expression of the nonclassical interference coefficient $\Lambda^\psi_{\bs\sigma,\bs\sigma',\bs\sigma''}$ [\cref{eq:Lambda}].
\\


We conclude this section by connecting the abstract discussion above to an operational strategy to simulate interference phenomena. 
In Ref.~\onlinecite{encoding} we prove that the total phase vector of a system of classical oscillators can be expanded over the sum of all eigenvalues of  the graph, including both emergent contributions (isolated peaks) and non-emergent terms. 
However, only the emergent states are associated to a quantum Hilbert space.
To resolve the quantum information encoded in the system it is therefore required to maximize the spectral gap between emergent and random states.
This is achieved via an ad-hoc choice of the valencies of the elementary graphs comprising the network, as discussed in \cref{sec:QLbits}.
Conveniently, this preparation stage needs to be defined once for all for any circuit, as the graph spectra are preserved by unitary QL gates.
Let consider as an example the specific case of LC circuits, already introduced in \cref{sec:perspectives}.
It is shown in Ref.~\onlinecite{maffezzoni2015} that the oscillatory patterns of these circuits allow to build Kuramoto-type models, where the interactions between oscillators are defined from transconductance elements.
After building this network and allowing for its long-time synchronization, a readout of the phases of the oscillators phases is required.
As the voltage $V_j(t)$ of each LC oscillator is related to its angle via $V_j(t) \propto\cos\left[\theta_j(t)\right]$, classical phase readout in this setup can be implemented by means of voltage probes, such as oscilloscopes or lock-in amplifiers. 
Finally, QL measurement processes can be simulated by numerically postprocessing the classical phase vector. 
For example, conditional probablities from \cref{eq:P_psi_cond} can be calculated by projecting the phases of the oscillators onto the spectrum of a QL observable operator $\mc F$.  These nonclassical probabilities obey by construction the five axioms listed above, mimicking nonclassical interference phenomena.

\section{Conclusions}

In this paper, we built upon the general framework developed in Refs.~\onlinecite{scholes2023, scholes2024, product} to define and manipulate QL states.
In \cref{sec:QLbits}, we constructed a computational basis from the emergent states of many-body classical systems, whose correlations can be mapped onto synchronizing graphs. We introduced the Hilbert space for an arbitrary number of QL bits and explicitly derived the Bell states in two dimensions.
In \cref{sec:gates}, we proposed a framework for defining quantum gates applicable to QL states and provided concrete examples of their action on many-body classical resources.
\Cref{sec:perspectives} offers an overview of experimental platforms that could be used to implement and test this theory.
Finally, in \cref{sec:quantum_prob}, we discussed that a QL notion of interference can emerge within this framework by presenting a set of axioms for interpreting QL states and observables.
\\

Several open questions remain, presenting diverse opportunities for future research.

As firstly discussed in \cref{subsubsec:many}, the number of edges in the Cartesian product graph defined in \cref{eq:R_NQL}, along with the associated gate transformations in \cref{eq:Rg_circ}, increases exponentially with the number of QL bits. 
This exponential scaling imposes practical constraints on the system sizes that can be investigated with current computational and experimental capabilities. Nevertheless, we believe that moderately sized networks are sufficient to demonstrate the core principles of QL algorithms. For example, entangled states such as Bell and GHZ states can be effectively generated and analyzed using shallow-depth circuits. \cite{product}
In terms of scaling, to date high-performance classical simulations allow the implementation of systems with up to approximately $\NQL \lesssim 5-10$ QL bits and $\Ng\lesssim 10$. 
Simulations of a Kuramoto model with this system size are given, e.g., in Ref.~\onlinecite{skardal2018}. 


It can be insightful to directly relate our work to the rigorous approach to quantum non-Markovianity proposed in Refs.~\onlinecite{pollock2018,white2020,berk2023}.
In that framework a notion of conditional probabilities is defined, that accounts explicitly for the role of measurement devices on the statistics. 
An important feature of that formalism is that it provides an operational definition of non-Markovianity which, differently from other strategies, converges to the correct Kolmogorov conditions in the classical limit. \cite{pollock2018nm}

A broad class of computational approaches to open quantum systems and quantum chemistry involves a perspective somehow complementary to our approach.
In particular, quantum nonadiabatic systems are often mapped onto quasiclassical models to minimize computational complexity and inexpensively approximate long-time dynamics.
Accurate quasiclassical methods have been derived, by requiring that Kolmogorov probabilities [fulfilling  \cref{eq:CFTP}] obey by construction quantum detailed balance at long times. \cite{thermalization,ellipsoid,CHIMIA,GQME,mannouch2023,qubit}
It can be insightful to assess whether even more accurate quasiclassical approaches can be derived from a QL stochastic theory which approximates \cref{eq:QFTP} beyond the first classical term.

The scope and applicability of our approach could be broadened by extending the treatment of quantum gates to any unitary transformation, in particular to time evolution operators. 
From this perspective, one direction for future research is to map the dynamics of a unitary quantum subsystem onto a large network of classical oscillators and, for instance, to explore the effects of decoherence arising from interactions with an external dissipative environment.
Accurately reproducing quantum time evolution through classical dynamics remains a nontrivial challenge.
It requires to extend our approach such that of the algebra of Hilbert space generators is preserved, at least approximately. \cite{runeson2019,kapral2015} 
This strategy becomes particularly promising when coupled with the enforcement of quantum detailed balance at long times---a direction the authors have recently explored in the context of nonadiabatic quantum dynamics. \cite{thermalization,ellipsoid}

The extent to which our framework offers any advantage compared to quantum computing remains an open question, that we believe will be fully explored by testing further approximations of the exact mapping proposed here.
More specifically, our approach requires an exponentially large number of oscillators to encode the correlations of the Hilbert space of a system of QL bits. 
This high resource demand does not offer per se any advantage in regards to scaling. 
In Appendix A of our new paper, Ref.~\onlinecite{encoding}, we outline as a potential strategy to mitigate the high cost of our approach by  replacing full graph representations with their lower rank approximations. 
This can lead to a renormalization of both graph resources and corresponding gates, possible by truncating negligible parts of the spectrum.
In general, we believe that it can be worth challenging the trade-off between implementing genuine quantum resources with polynomial complexity and easier accessible classical resources with unfavorable exponential scaling.
We view these observations as a starting point for further investigations into the foundational limits and capabilities of our approach and quantum-inspired frameworks in general.

Future work could explore new possibilities regarding the choice of resources used for information processing.
For example, our framework could be extended to other flavors of Kuramoto-type models, where oscillators are represented by vectorial variables \cite{amdeaguiar2023} or elements of a unitary group. \cite{lohe2009,scholes2022}
Other classical systems could be used as a resource for information processing in a similar fashion as what we proposed here.
An example in this regard is the Fermi--Pasta--Ulam--Tsingau chain, where isolated solitary waves, emerging from symplectic classical nonlinear dynamics, could be used to store and process information in a QL fashion. \cite{memoryFPU,FPUglasses,quantumFPU,berman2005} 
Since the dynamics of this classical model are Hamiltonian, the approach may allow for a well-defined classical detailed balance condition, which could be enforced to match its quantum analog and ensure correct long-time thermalization. \cite{thermalization}

Finally, it can be a fascinating topic for future research to connect QL information processing introduced here to other information theories on human decision-making and psychology derived in the framework of QL contextual probability models. 
\cite{khrennikov2016,ozawa2020,scholes2024,widdows2023}
Implementing our formalism on brain-like neuromorphic circuits \cite{pal2024} can support theoretical and experimental investigations in this framework.

\section{Acknowledgments}

The authors thank Dr. Debadrita Saha for useful discussions. This research was funded by the National Science Foundation under Grant No. 2211326 and the Gordon and Betty Moore Foundation through Grant GBMF7114.

\begin{appendix}
	
\section{General transformation from emergent states to the computational basis}\label{app:Ucb}

In \cref{sec:gates} we discussed how a QL representation of quantum gates can be straightforwardly defined once the emergent states are mapped onto the computational basis.
In general, this transformation cannot be expressed in analytic form as soon as the subgraphs defining the resources are not exactly regular. 
However, we show here that the problem of determining the map can be rewritten as a simple matrix inversion problem.

Firstly, it is convenient to parametrize the generalization of \cref{eq:Ucb} as follows:
\begin{equation}\label{eq:Ucb_blocks}
U_{\mr{cb}} =\bigotimes_{q=1}^\NQL \begin{pmatrix}  X^{(q)} & Y^{(q)}  \\ W^{(q)}  & Z^{(q)} \end{pmatrix}, 
\end{equation}
where $X^{(q)} \in \mathbb R^{\Ng}$ is diagonal, with components $X_{ij}^{(q)}  =\delta_{ij}x_j^{(q)} $, and similarly for the other three blocks in \cref{eq:Ucb_blocks}. 
\Cref{eq:Ucb_blocks,eq:Ucb_gen} can be rewritten as the linear problem
\begin{subequations}\label{eq:Ucb_gen}
\begin{align}
x_i^{(q)} a_{1,i}^{(q)} + y_i^{(q)} a_{2,i}^{(q)} &= \Ng^{-1/2},\\
w_i^{(q)} a_{1,i}^{(q)} + z_i^{(q)} a_{2,i}^{(q)} &= 0, \\
x_i^{(q)} a_{1,i}^{(q)} - y_i^{(q)} a_{2,i}^{(q)} &= 0,\\
w_i^{(q)} a_{1,i}^{(q)} - z_i^{(q)} a_{2,i}^{(q)} &= \Ng^{-1/2},
\end{align}
\end{subequations}
where $q=1,\ldots,\NQL$, $i=1,\ldots, \Ng$.
\Cref{eq:Ucb_gen} is expressed in matrix form as 
\begin{align}\label{eq:lin_gate}
&\begin{pmatrix}
D_1^{(q)} & D_2^{(q)} & \mathbb 0_\Ng & \mathbb 0_\Ng \\
\mathbb 0_\Ng & \mathbb 0_\Ng & D_1^{(q)} & D_2^{(q)} \\
D_1^{(q)} & -D_2^{(q)} & \mathbb 0_\Ng & \mathbb 0_\Ng \\
\mathbb 0_\Ng & \mathbb 0_\Ng & D_1^{(q)} & -D_2^{(q)} 
\end{pmatrix}
\begin{pmatrix}
x^{(q)} \\ y^{(q)}\\ w^{(q)}\\ z^{(q)} 
\end{pmatrix} =\begin{pmatrix}
1_\Ng \\ 0_\Ng\\ 0_\Ng\\ 1_\Ng 
\end{pmatrix},
\end{align}
where $\left[D_n^{(q)}\right]_{ij} = \delta_{ij} a_{n,j}^{(q)}$, $n=1,2$. 
\Cref{eq:lin_gate} can be inverted to determine a solution for $\left(x^{(q)},y^{(q)},w^{(q)},z^{(q)}\right)$ $\forall q=1,\ldots \NQL$, corresponding to a unique expression for $U_{\mr{cb}}$.

\section{Derivation of the quantum-like formula of total probabilities }\label{app:QFTP}

In this appendix we derive \cref{eq:QFTP}, describing the update of probabilities for two QL observables $\mc F$ and $\mc G$, with eigenvalue problems defined as in \cref{eq:eigFG}. 

Any state vector $\psi \in \mc H$ can be expanded as
\begin{equation}\label{eq:psi_FG}
	\psi = \sum_{\bs \sigma\in \bs{\mc S}} G^\psi_{\bs \sigma} g_{\bs \sigma}= \sum_{\bs \sigma'\in \bs{\mc S}} F^\psi_{\bs \sigma'} f_{\bs \sigma'} ,
\end{equation}
where $F^\psi_{\bs \sigma'} = \braket{f_{\bs \sigma'}, \psi}$ and $G^\psi_{\bs \sigma} = \braket{g_{\bs \sigma}, \psi}$.
We can connect the probabilities of the two observables by expressing
\begin{subequations}\label{eq:GF_gen}
\begin{align}
	\psi &= \sum_{\bs\sigma,\bs \sigma' \in\bs{\mc S}}F^\psi_{\bs \sigma'} u_{\bs \sigma',\bs \sigma}g_{\bs \sigma},\\
	u_{\bs \sigma',\bs \sigma} &= \braket{g_{\bs \sigma},f_{\bs \sigma'}},
\end{align}
\end{subequations}
which leads to the following mapping between the coefficients of $\psi$ in the two bases:
\begin{equation}\label{eq:GF}
	G^\psi_{\bs \sigma} = \sum_{\bs\sigma'\in \bs{\mc S}} F^\psi_{\bs \sigma'}u_{\bs\sigma',\bs\sigma}.
\end{equation} 
To relate \cref{eq:GF} to QL probabilities, we define
\begin{equation}
	F^\psi_{\bs \sigma'} = \e^{i \gamma_{\bs \sigma'}^\psi} \sqrt{ P_\psi\left(\lambda_{\bs \sigma'}^{\mc F}\right)}, \hspace{4mm}
	u_{\bs\sigma',\bs\sigma} = \e^{i \Gamma_{\bs \sigma',\bs\sigma}}\sqrt{P\left(\lambda^{\mc F}_{\bs \sigma'}|\lambda^{\mc G}_{\bs \sigma}\right)},
\end{equation}
where $\gamma^\psi_{\bs \sigma'}, \Gamma_{\bs \sigma',\bs \sigma}\in [0,2\pi)$.
By expanding the squared modulus of \cref{eq:GF}, we finally obtain \cref{eq:QFTP}. 
This involves the phase factor
\begin{equation}\label{eq:Lambda}
\Lambda^\psi_{\bs\sigma,\bs\sigma',\bs\sigma''} = \e^{i\left(\gamma^\psi_{\bs\sigma'} - \gamma^	\psi_{\bs\sigma''} \right)}\e^{i\left(\Gamma_{\bs\sigma',\bs\sigma}- \Gamma_{\bs\sigma'',\bs\sigma}\right)},
\end{equation}
which, being any other term of \cref{eq:QFTP} real, can be replaced by its real part:
\begin{equation}\label{eq:Lambda_cos}
\Lambda^\psi_{\bs\sigma,\bs\sigma',\bs\sigma''} \mapsto \cos \left[(\gamma^\psi_{\bs\sigma'} - \gamma^\psi_{\bs\sigma''})(\Gamma_{\bs\sigma',\bs\sigma}- \Gamma_{\bs\sigma'',\bs\sigma})\right].
\end{equation}

\section{Spectral properties of QL observables}\label{app:spectrum}
In this appendix we study the spectral properties of QL observables [\cref{eq:F}], that are Hermitian operators according to the definition \cref{eq:Psi_dagger}.
In particular, we show that \ref{enum:i} their eigenvalues are real, \ref{enum:ii} eigenvectors corresponding to different eigenvalues are orthogonal, and \ref{enum:iii} there exists an orthonormal basis of $\mc H$ consisting of eigenvectors.
We refer here to the eigenvalues and eigenvectors of $\mc F$ as $\lambda_{\bs\sigma}$'s and $f_{\bs \sigma}$'s, respectively. 

\begin{enumerate}[label=(\roman*)]
	\item\label{enum:i} For an arbitrary $f_{\bs \sigma}$ 
	\begin{align*}
		\braket {f_{\bs\sigma}, \mc F f_{\bs\sigma}} = \lambda_{\bs\sigma}\norm{f_{\bs\sigma}}^2= \braket{  \mc Ff_{\bs\sigma}, f_{\bs\sigma}} = \lambda_{\bs \sigma}^* \norm{f_{\bs\sigma}}^2,
	\end{align*}
which implies that $\lambda_{\bs \sigma}\in \mathbb R$.
	\item\label{enum:ii} Suppose that $\lambda_{\bs\sigma}\neq \lambda_{\bs\sigma'}$ for $\bs\sigma \neq \bs\sigma'$. By multiplying 
	\begin{equation}\label{eq:eig_ii}
		\mc Ff_{\bs \sigma} = \lambda_{\bs \sigma}f_{\bs \sigma},
	\end{equation}
 by $f_{\bs\sigma'}$, we obtain
	\begin{align}
		\langle f_{\bs\sigma'}, \mc Ff_{\bs\sigma}\rangle &= \lambda_{\bs \sigma}\langle f_{\bs\sigma'}, f_{\bs\sigma}\rangle\nn\\
		&=\langle \mc F f_{\bs\sigma'}, f_{\bs\sigma}\rangle \substack{\ref{enum:i}\\=} \lambda_{\bs \sigma'}\langle f_{\bs\sigma'}, f_{\bs\sigma}\rangle,
	\end{align}
	which implies that $\langle f_{\bs\sigma'}, f_{\bs\sigma}\rangle=0$.
	\item \label{enum:iii} 
	Let us order the eigenvalues and eigenvectors of $\mc F$ in increasing order, via the map
	\begin{equation} \label{eq:sigma_to_n} 
		\bs \sigma \mapsto n = \sum_{l=0}^{\NQL-1}2^l\delta_{\sigma_l,\ua}.
	\end{equation}
	We denote by $V_1$ the set of vectors orthogonal to the eigenvector $f_1$  of $\mc F$ with eigenvalue $\lambda_1$. 
	Note that $\mc F$ maps $V_1$ onto itself, i.e., $\mc F \psi \in  V_1$ for $\psi\in V_1$. In fact,
	\begin{equation}
		\braket{f_1,\mc F\psi} = \braket{\mc F f_1,\psi} =\lambda_1 \braket{f_1,\psi} = 0.
	\end{equation}
	The linear operator $L(\phi) = \mc F \phi$, when restricted to $V_1$, is also Hermitian, and it admits an eigenvalue $\lambda_2$ with eigenvector $f_2\in V_1$. 
	By construction, $f_2$ is orthogonal to $f_1$. 
	We then define by $V_2$ the orthogonal complement to $\mr{span}\{f_1,f_2\}$.
	Again, $\mc F$ maps $V_2$ onto itself.
	By proceeding this way, we find a sequence $\{\lambda_n, f_n,V_n\}_{n}$, where the subspace $V_n$ is orthogonal to  $\mr{span}\{f_1,\ldots, f_n\}$.
	The sequence ends at the step of order $2^\NQL$, given that $\dim(V_n) = 2^\NQL -n$. 
	This process allows us to define a complete set of mutually orthogonal eigenvectors.
\end{enumerate}
\vspace{0.2cm}

\section{Dynamics of 
 the Kuramoto model} \label{app:kuramoto}

Here, we present a closed-form solution for the dynamics of the Kuramoto model, as described by the nonlinear equations of motion in \cref{eq:kuramoto}.
Our approach follows the methods outlined in Refs.~\onlinecite{muller2021,kawamura2010}. 
This analysis offers a practical framework for implementing the dynamics of the model efficiently, without the need for explicitly integrating the equations of motion, making it applicable to a broad range of synchronizing networks. Additionally, this work can serve as a starting point for future analytical studies exploring the role of emergent states in synchronization processes.

If the coupling matrix $\mc R$ is real, it is convenient to subtract from \cref{eq:kuramoto} an additional complex component,
\begin{align}\label{eq:kuramoto_compl}
\dot \theta_i &= \omega_i +\frac{\Gamma}{2\Ng}\sum_{j=1}^{2\Ng} \mc R_{ij}\left[\sin(\theta_j-\theta_i)-i\cos(\theta_j-\theta_i) \right], 
\end{align}
 leading to
\begin{equation}\label{eq:dot_ith}
i\dot \theta_i = i\omega_i +\frac{\Gamma}{2\Ng}\e^{-i\theta_i}\sum_{j=1}^\NQL \mc R_{ij}\e^{i\theta_j}.
\end{equation}
\Cref{eq:dot_ith} can be expressed in matrix form as
\begin{align}
&\frac{\mr d}{\mr d t} \e^{i\bs \theta} = \left[\diag (i\bs\omega)+\frac{\Gamma}{2\Ng}\mc R\right] \e^{i\bs \theta}
\end{align}
or, more compactly as, $\dot{\bs x} = \tilde{\mc R}\bs x$, where $\bs x = \e^{i\bs\theta} = (\e^{i\theta_1},\ldots, \e^{i\theta_{2\Ng}})^\top$ and $\tilde{\mc R} = \mr{diag}(i\bs\omega)+\frac{\Gamma}{2\Ng}\mc R $.
This allows us to obtain as a closed form for the solution of the dynamics of the Kuramoto model
\begin{equation}\label{eq:xt_kur}
\bs x(t) =  \e^{\tilde{\mc R} t}\bs x(0).
\end{equation}
We can decompose the real and imaginary part of the frequency vector $\bs \theta = \bs \theta_{\mr{re}}+i \bs \theta_{\mr{im}}$, to obtain as a physical solution of \cref{eq:kuramoto_compl}
\begin{align}
i\bs \theta_{\mr{re}}(t)-\bs\theta_{\mr{im}}(t)
&=  \log \left[ \e^{\tilde{\mc R} t}\bs x(0)\right],
\end{align}
or 
\begin{subequations}\label{eq:sol_theta_alg}
\begin{align}
\bs \theta_{\mr{re}}(t)&=  \Im\left\{\log \left[ \e^{\tilde{\mc R} t}\bs x(0)\right]\right\},\\
\bs \theta_{\mr{im}}(t)&=  -\Re\left\{\log \left[ \e^{\tilde{\mc R} t}\bs x(0)\right]\right\}.
\end{align}
\end{subequations}

\end{appendix}

\FloatBarrier
%

\end{document}